\documentclass[AMA]{WileyASNA-v1}

\articletype{Original paper}

\received{}
\revised{}
\accepted{}

\usepackage{caption}
\usepackage{amsmath}

\begin{document}
\newcommand{\farcs}{$.\!\!^{\prime\prime}$}
\newcommand{\arcsec}{$^{\prime\prime}$}

\newcommand{\arcmin}{$^{\prime}$}
\title{The isolated elliptical galaxy NGC 5812 - MOND or Dark Matter? 
%\title{Globular cluster systems of isolated elliptical galaxies.  IV: NGC 5812 - another  dark matter poor galaxy 
\protect\thanks{Based on observations obtained with Gemini-South,  Cerro Pachon, Chile, under the programme GS-2013A-Q-051.}}

\author[1]{Tom Richtler}

\address[1]{\orgdiv{Departamento de Astronom\'{\i}a}, \orgname{Universidad de Concepci\'on, Concepc\'ion},  \orgaddress{\state{} \country{Chile}}}
	
\author[2]{Ricardo Salinas}

\address[2]{\orgdiv{Departamento de Astronom\'{\i}a, Universidad de La Serena, Av. Juan Cisternas 1200 Norte, La Serena},  \orgaddress{\state{} \country{Chile}}}
           
\author[3]{Richard Lane}
\authormark{Richtler \textsc{et al}}
%Centro de Investigaci\'on en Astronom\'ia, Universidad Bernardo O'Higgins, Avenida Viel 1497, Santiago, Chile"
\address[3]{\orgdiv{Centro de Investigaci\'on en Astronom\'ia, Universidad Bernardo O'Higgins, Avenida Viel 1497, Santiago},  \orgname{} \orgaddress{\state{} \country{Chile}}}

\author[4]{Michael Hilker}

\address[4]{\orgdiv{} \orgname{ European Southern Observatory, Karl-Schwarzschildstr. 2,  85748, Garching},    \orgaddress{\state{} \country{Germany}}}
        
             % \and 
	%European Southern Observatory, Karl-Schwarzschild-Str.~2,
            %    D-85748 Garching, Germany               

\corres{Tom Richtler \email{tom@astro-udec.cl}}
%\presentaddress{}
\abstract{%The case for a MONDian phenomenology also among early-type galaxies  has become so strong that deviations from a
%MONDian behaviour appear more interesting than agreements.
%Their dark halos (if they exist) should be largely unaffected by galaxy interaction. However, 
There exist  isolated elliptical galaxies, whose dynamics can be modelled without resorting to dark matter or MOND, e.g. NGC 7507.
Such objects lack
understanding within the current framework of galaxy formation.
 %It is therefore of high interest to find and understand more examples.  }
The isolated elliptical NGC 5812  is another object to investigate a possible role of isolation. We use globular clusters (GCs) and the galaxy light itself as dynamical tracers to constrain its mass profile.
We  employ  Gemini/GMOS mask spectroscopy, apply the GMOS reduction procedures provided within IRAF,  measure  GC velocities  by cross correlation methods and extract  the line-of-sight kinematics of galaxy spectra using the tool pPXF.
We identify 28 GCs with an outermost galactocentric distance of 20 kpc, for which velocities could be obtained.  Furthermore, 16 spectra of the integrated galaxy light  out to 6 kpc have been used to model the central kinematics. These spectra provide
evidence for a disturbed velocity field, which is plausible given the disturbed morphology of the galaxy.
  We construct spherical Jeans models for the galaxy light and apply tracer mass estimators for the globular clusters. 
With the assumptions inherent to the mass estimators,  MOND is compatible with the  mass  out to   20 kpc. However,  a dark matter free galaxy is not excluded, given the uncertainties related to a possible non-sphericity  and a possible non-equilibrium state. 
   We find one globular cluster with an estimated mass of $1.6\times10^7 M_\odot$, the first Ultra Compact Dwarf in an isolated elliptical.
% With NGC 5812 we have found another  example of an isolated elliptical, where dark matter is not directly visible. 
 We put NGC 5812 into the general context of dark matter or alternative ideas in elliptical galaxies.   
The case for a MONDian phenomenology also among early-type galaxies  has become so strong that deviating cases 
appear astrophysically more interesting than agreements. The baryonic Tully Fisher relation (BTFR) as predicted by MOND is observed in some samples of early-type galaxies, in others not.
However, in cases of  galaxies that deviate from the MONDian prediction, data quality and data completeness  are often problematic. 
%Understanding, why galaxies like NGC 5812 and NGC 7507 deviate from the BTFR
%might  be critical for the acceptance of a universal BTFR. 
} 
%Our result is not a cheap falsification of fundamental physics. 

\keywords{Galaxies: individual: NGC 5812 -- Galaxies: kinematics and dynamics -- Galaxies: star clusters}

\maketitle

\section{Introduction}

"Galaxies are embedded in halos of dark matter (DM) that assemble through cosmological structure formation." This statement is the core of the present paradigm
of galaxy formation in a "concordant" universe \citep{ostriker95}.   The most influential description of this  concordant universe  is found in  the $\Lambda$CDM cosmological parameters from the 
Planck satellite analyses \citep{planckcoll14},
where the baryonic density parameter $\Omega_{baryon}$ is 0.04 and the matter density parameter   $\Omega_{matter}$ is 0.3 (uncertainties and model dependencies are omitted for simplicity),
 if a Friedmann cosmological model is assumed.  Therefore, under this paradigm dark matter (DM)  plausibly consists of 
 non-baryonic particles \citep{bertone05,bertone10}. Their non-detections can be understood as a crisis in the understanding  of the DM concept \citep{bertone18a}.
However, dark matter particles might be detected in future   terrestrial experiments or  by observing  DM annihilation 
 products \citep{silverwood15, acharya19}. See also  \citet{bertone18b} for an historical account of the dark matter concept. 
 Currently, the tension between the locally measured Hubble constant and the Hubble constant derived from Planck data, as well as other tensions between cosmological data sets, cast doubt on the very
 Friedmann concept of a homogeneous universe (e.g. \citealt{divalentino20}, \citealt{heinesen20}). 

This "standard cosmology" has never been accepted by the entire community.  A series of papers   \citep{milgrom83a,milgrom83b,milgrom83c,bekenstein84} marked the beginning
of an alternative view onto the mass discrepancy which became known as Modified Newtonian Dynamics (MOND).  
%While cosmological simulations at high  and low redshifts are successful in reproducing the cosmic web, 
Dynamical features in local
galaxies   are difficult to reconcile with the nature of dark matter as weakly interacting massive particles \citep{famaey12,famaey13,kroupa15}. Of particular interest is the (predicted by Milgrom) intimate
relation between the baryonic matter and dark matter that has become known as the MONDian phenomenology manifest in rotation curves and in the baryonic Tully-Fisher relation of spiral 
galaxies \citep{famaey12,famaey13,kroupa15,mcgaugh16,lelli19}.

The knowledge regarding the dynamics of {\it early-type galaxies} is naturally  poorer because of the missing disk symmetry. 
Using globular clusters and planetary nebulae as dynamical tracers at larger radii as well as the hydrostatics of the hot halo gas, 
some elliptical galaxies have been shown to be  consistent with MONDian behaviour  \citep{weijmans08, schuberth12, milgrom12, lelli17}. 
 However, central galaxies in galaxy clusters 
apparently need an additional dark component to explain their kinematics \citep{richtler08,richtler11} as it is the case with entire galaxy clusters \citep{sanders03,pointecouteau05,angus08}
(but see the remarks in Section \ref{sec:dmdomination}).

There is a class of early-type galaxies showing a radial behaviour of the velocity dispersion that has been dubbed "Keplerian decline" (it is  of course not a Keplerian potential),
meaning that models can work without a dark component \citep{romanowsky03}. 
One of Romanowsky et al.'s  prominent objects is NGC 3379, which one may rank among the dynamically best known elliptical galaxies. \citet{delorenzi09}  demonstrate that the available data are
  consistent with a variety of dynamical models and dark matter abundances. However, to permit a cosmologically motivated dark halo, a very strong radial anisotropy of the galaxy's stellar orbits is demanded, which so far has never been observed. In this context, isolated ellipticals are particularly interesting because the assembly history of their dark halos
  was not disturbed by a cluster or group environment (see \citealt{richtler15} for a brief introduction into the literature on isolated ellipticals). As an example, we mention
NGC 7507 as an  even more convincing galaxy with  a "Keplerian"  behaviour of the velocity dispersion of the integrated galaxy light  \citep{salinas12,lane15}.

%The SLUGGS survey provided a lot of globular cluster velocities.

In this paper we show and interpret kinematical data for another isolated elliptical, NGC 5812. The globular cluster system and the galaxy's morphology in the Washington system has been studied by \citet{lane13}. NGC 5812 is, judging from its photometric properties, an intermediate-age early-type galaxy, showing deviations from a smooth light distribution that plausibly are the relics of previous infall/interaction processes.
A companion with a tidal tail is the obvious manifestation of a present galaxy-galaxy interaction.

The heliocentric radial velocity of NGC 5812 is 1970 km/s (NED).
 We adopt a distance of 28 Mpc (m-M = 32.24) \citep{lane13}. At this distance, 1''
 corresponds to 135.7 pc.

 \section{Observations and Data Reduction}
\subsection{Observations}

The observations  were performed in queue mode during  the period May 6th to August 6th 2013
 at the Gemini South Observatory 
at Cerro Paranal, Chile (programme GS-2013A-Q-51, PI:Richtler). The  Gemini Multi-Object Spectrograph South (GMOS-S)  was used.
The disperser was the B600+\_G532 grism,  giving a resolution of $\sim$4.7\AA~FWHM. 
The detector was a mosaic of three CCDs. In 2$\times$ binned mode, the pixel scale is 0.15\arcsec and the total field of view  5.5\arcmin$\times$ 5.5\arcmin. 

Three spectroscopic masks were exposed, two of them centred on NGC 5812, while  the third mask was shifted by 2.73\arcmin to the
North.  The observations are summarized in Table \ref{tab:obs}. Table entries are the mask number, the coordinates, exposure time, seeing, total number of slits, Gemini exposure coding
(which gives the date), and the Universal Time of the exposure start.

%In the MXU mode, up to ten masks can be loaded into the magazine of
%the Mask EXchange Unit (MXU) during daytime. The selected mask is
%moved in and out of the focal plane by a drive. 

%For objects located close to the edge of the field-of-view (in
%dispersion direction), a part of the spectrum will not be projected
%onto the CCD: for slits near the left (right) border, the blue (red)
%part of the spectrum will be truncated. The mask layout is done by the
%observers prior to observations. This crucial step is described in the
%following section. DISPERSION!

\begin{table*}
\centering
\begin{tabular}{cllrcrcr}\hline
\hline
{Mask} & \multicolumn{2}{c}{Center Position} &
{Exp.~Time}&{Seeing}&{\#\,Slits}&{OB Id}&{UT} \\
{} & \multicolumn{2}{c}{{({J\,2000})}} & {{({sec})}} & &  & &{(start)\,} \\
\hline
%\hline

 1& 15:00:55 & -7:27:26 & 1400 &  0\farcs 9 & 37 &S20130506S0089 & 6:07 \\   
 1& 15:00:55 & -7:27:26 & 1400 &  0\farcs9& 37 &S20130506S0092 & 6:34 \\ 
 1& 15:00:55 & -7:27:26 & 1400 &  0\farcs 8& 37 &S20130506S0095 & 7:00 \\
 
 1& 15:00:55 & -7:27:26 & 1400 &  0\farcs 8& 37 &S20130603S0078 & 3:16 \\
 1& 15:00:55 & -7:27:26 & 1400 &  0\farcs 8& 37 &S20130603S0081 & 3:42 \\
 1& 15:00:55 & -7:27:26 & 1400 &  0\farcs 8& 37 &S20130603S0084 & 4:09 \\

  2& 15:00:55 & -7:27:26 & 1430 &  0\farcs 9& 47 &S20130605S0138 & 0:39 \\
  2& 15:00:55 & -7:27:26 & 1430 &  0\farcs 8& 47 &S20130605S0144 & 5:12 \\  
  2& 15:00:55 & -7:27:26 & 1400 &  0\farcs 9& 47 &S20130807S0025 & 0:13 \\
  
  2& 15:00:55 & -7:27:26 & 1400 &  1\farcs 1& 47 & S20130807S0028 & 0:13 \\
 2& 15:00:55 & -7:27:26 & 1400 &  1\farcs 2& 47 & S20130807S0031 & 1:06 \\
 2& 15:00:55 & -7:27:26 & 1400 &  1\farcs 0& 47 & S20130810S0038 & 0:21 \\

  3& 15:00:55 & -7:24:10 & 1070 &  0\farcs 8& 22 & S20130702S0072 & 2:36 \\
  3& 15:00:55 & -7:24:10 & 1070 &  0\farcs 9& 22 &S20130702S0075 & 2:58 \\     
  3& 15:00:55 & -7:24:10 & 1070 &  1\farcs 1& 22 &S20130702S0078& 3:19\\   
  3& 15:00:55 & -7:24:10 & 1500 &  0\farcs 9& 22 &S20130806S0075& 1:10\\ 
  3& 15:00:55 & -7:24:10 & 1500 &  0\farcs 8& 22 &S20130806S0080& 1:41\\ 
    3& 15:00:55 & -7:24:10 & 1500 &  0\farcs 6& 22 &S20130806S0082 & 2:12\\   
   \hline
\hline
\end{tabular}
\caption{List of mask characteristics.}
\label{tab:obs}
\end{table*}

\subsection{Mask preparation}
\label{sect:masks}
\begin{figure*}[t]
\centering
\includegraphics[width=0.8\textwidth]{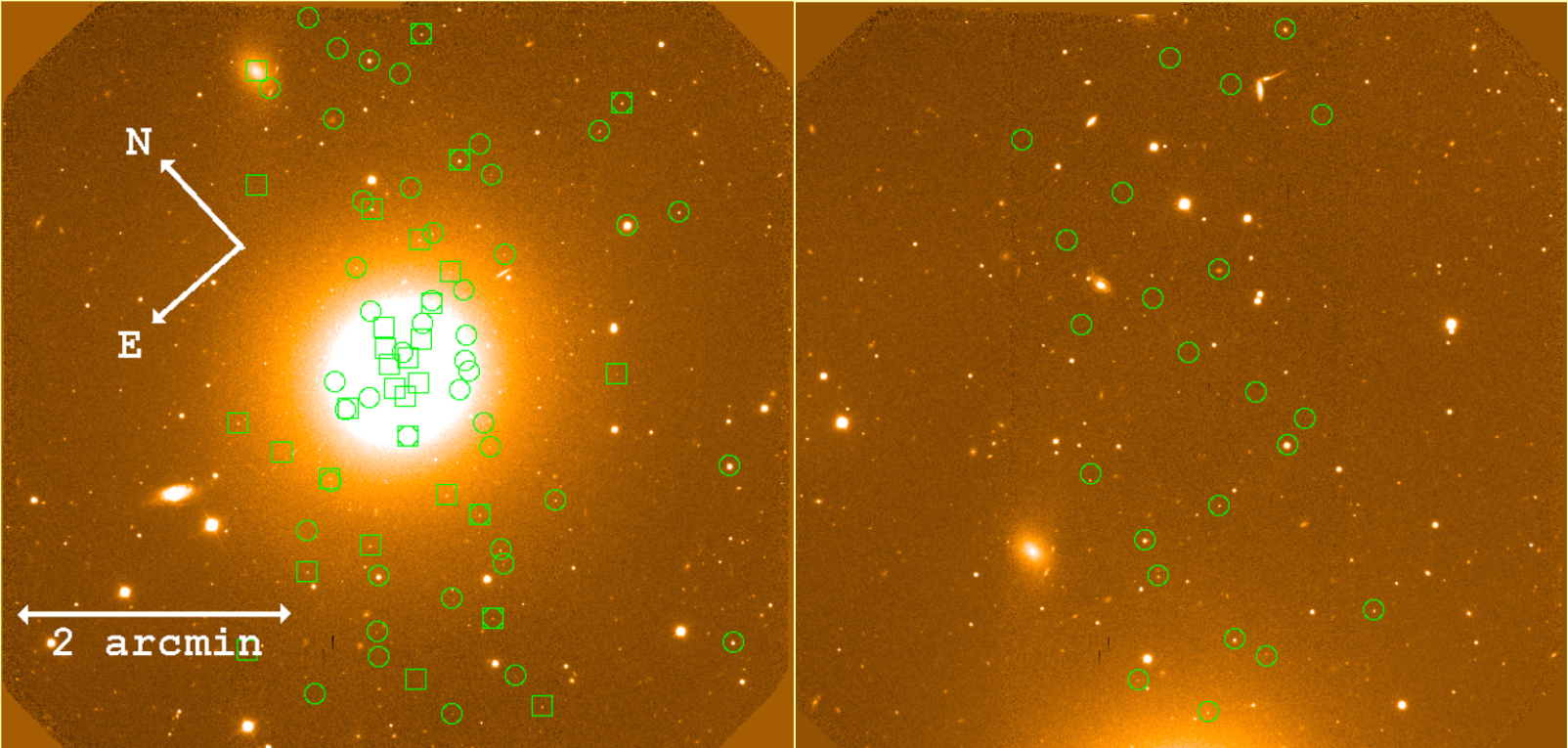}
%\centerline{\resizebox{\hsize}{!}{\includegraphics[angle=0]{allmasks.pdf}}} 
\caption{Left panel: Slit positions for mask 1 (squares) and mask 2 (circles), overlaid on a GEMINI preimage for this run. The brightest
globular cluster is the bright object  1.5\arcmin north-west of the centre. Right panel:
slit positions for mask 3, overlaid on a Gemini preimage}
\label{fig:slits}
\end{figure*}

%A very important step in the investigation was the preparation of the
%MXU slit masks.
Preimaging  of the three  fields (see Fig.\ref{fig:slits}) was carried out on February 2013. Each field was
observed in the r\_G0326-filter for 30.5 seconds. For the candidate
selection we used the  photometry  in the Washington system by 
\citet{lane13}. 
  Cluster candidates had to fulfill the following
criteria: the allowed color range was $0.9 < C-R < 2.1$, which is the appropriate color range for old GCs (e.g. \citealt{dirsch03}). Moreover, the
candidates should exhibit a star-like appearance on the pre-images to
distinguish them from background galaxies.  
%At the time of mask preparations, we
%were not aware of the possible existence of younger populations. The mean age of NGC 5812 is about 5 Gyr  so we may 
%picked up  younger  GCs only by chance.
The slit lengths were chosen  
 by the contrasting demands to sample as many GC candidates as possible (short slits), but also allow to measure
 the galaxy light  and the sky with sufficient accuracy (long slits).  In total we set  107 slits with lengths between 2\arcsec and 15 \arcsec .
 All slits have the same width of 1\arcsec. 

The mask design has been performed with the GEMINI software gmmps.

%\begin{description}
%\item[bias:]
%\item[flat fields:]
%\item[wavelength calibrations:]
%\end{description}

\subsection{Reductions of the GMOS masks and radial velocity measurements} 

The tools for the reduction were taken  from the GMOS package of IRAF. 
%In the appendix we give a
%typical reduction session, using as example  the first exposure from table \ref{tab:observations}.
%There is an  associated flat field exposure   
%S20130506S0090, and an an  associated arc exposure,S20130506S0091. 

 For the meaning of the IRAF/GMOS commands, we refer the reader to the IRAF manual \footnote{www.iraf.net}.
 The result is a multi-extension fits file with each slit being an extension.

The   reduction procedure and the measurement of radial velocities  has been already described in numerous
other papers, e.g. \citet{richtler04, schuberth06, schuberth10,schuberth12}, and here we provide a short summary.

For basic reduction, spectrum extraction and  wave-length calibration, we used the IRAF-task {\it identify} and {\it apall}. 
%In total, we extracted and calibrated  about 600 spectra, of which 178 were GCs, ?? stars, 16 quasars, and ?? galaxies. 

The radial velocities have been determined
using the cross-correlation IRAF-task {\it fxcor}.  Due to the very different appearance and S/N of the spectra, it turned out to
be impossible to establish a standard procedure, which would always use the same task parameters. Regarding the cross-correlation
interval, we found the interval 4700\AA -5400 \AA  to produce the best results.  Clearly defined correlation peaks are connected with
uncertainties around 20-30 km/s.   In the case of faint sources, more than one peak might appear, depending on the exact wavelength
interval, within which the correlation is done. In these cases, we tested which peak was the most stable against variations of the cross-correlation interval.  
%The uncertainty then may not be the uncertainty suggested by the
%broadness of the correlation peak.
 We used as templates a high S/N spectrum of NGC 1396, obtained with the same instrumentation during an earlier run \citep{richtler04} and a spectrum of one of the brightest globular clusters in NGC 4636 \citep{schuberth10}.

The globular cluster data  are  presented in  Appendix A.
% \citep{kumar12}.

%Fig.\ref{fig:uncertainties} shows the velocity uncertainties in dependence on the R-magnitude. The uncertainties are directly taken
%from {\it fxcor}. They  cluster around 50 km/s as in previous work. 

%\begin{figure}[]
%\begin{center}
%\includegraphics[width=0.5\textwidth]{fehler_5812.pdf}
%\caption{Uncertainties of the radial velocities in dependence on the R-magnitude. }
%\label{fig:uncertainties}
%\end{center}
%\end{figure}

\section{Radial velocities of globular clusters}
% and the analysis of the galaxy light}

\subsection{The velocity sample }

The tables  A1, A2 and A3 (appendix)  list all measured velocities  for the galaxy light, for  29 GCs and 19 stars, respectively. The velocities derived from the two templates are listed differently
to mediate an impression of the template's significance. 
%In a few cases of low S/N, the GC template does not reveal a trustworthy result.   

\subsection{Systemic velocity}
The systemic velocity of NGC 5812 is 1970 km/s, according to the NED. Defining our systemic velocity as the mean of the galaxy velocities and
the globular cluster velocities, we end up with a somewhat lower value.  The mean value and its mean uncertainty of the globular cluster velocities is
1922$\pm$22 km/s (from table  A2) .The galaxy velocities have a mean value of 1927$\pm$7 km/s (from table A1), which is excellent agreement.
% given the fact the galaxy velocities
%have been measured with different templates. 
 We therefore adopt 1925 km/s as the systemic velocity of NGC 5812. The reason for the difference to
the NED values remains obscure.

%The velocities of the integrated galaxy light have been measured using both templates.  The difference velocity(template:NGC1396) - velocity(template:GC) is +12$\pm$17 km/s.
%We constructed the final velocities accordingly by adding 12 km/s to the velocities from the GC-template and took the mean values.

\subsection{Velocity distribution of globular clusters and estimation of virial mass}
The standard deviation of the entire GC sample is 129$\pm$km/s, including the outlier at 1510 km /s (object 128 in Table A2). Without it the standard deviation shrinks to 109 km/s. 
The maximum likelihood dispersion estimator of \citet{pryor93} that also respects the individual uncertainties, gives 91$\pm$14 km/s. 
 
\begin{figure}[]
\begin{center}
\includegraphics[width=0.4\textwidth]{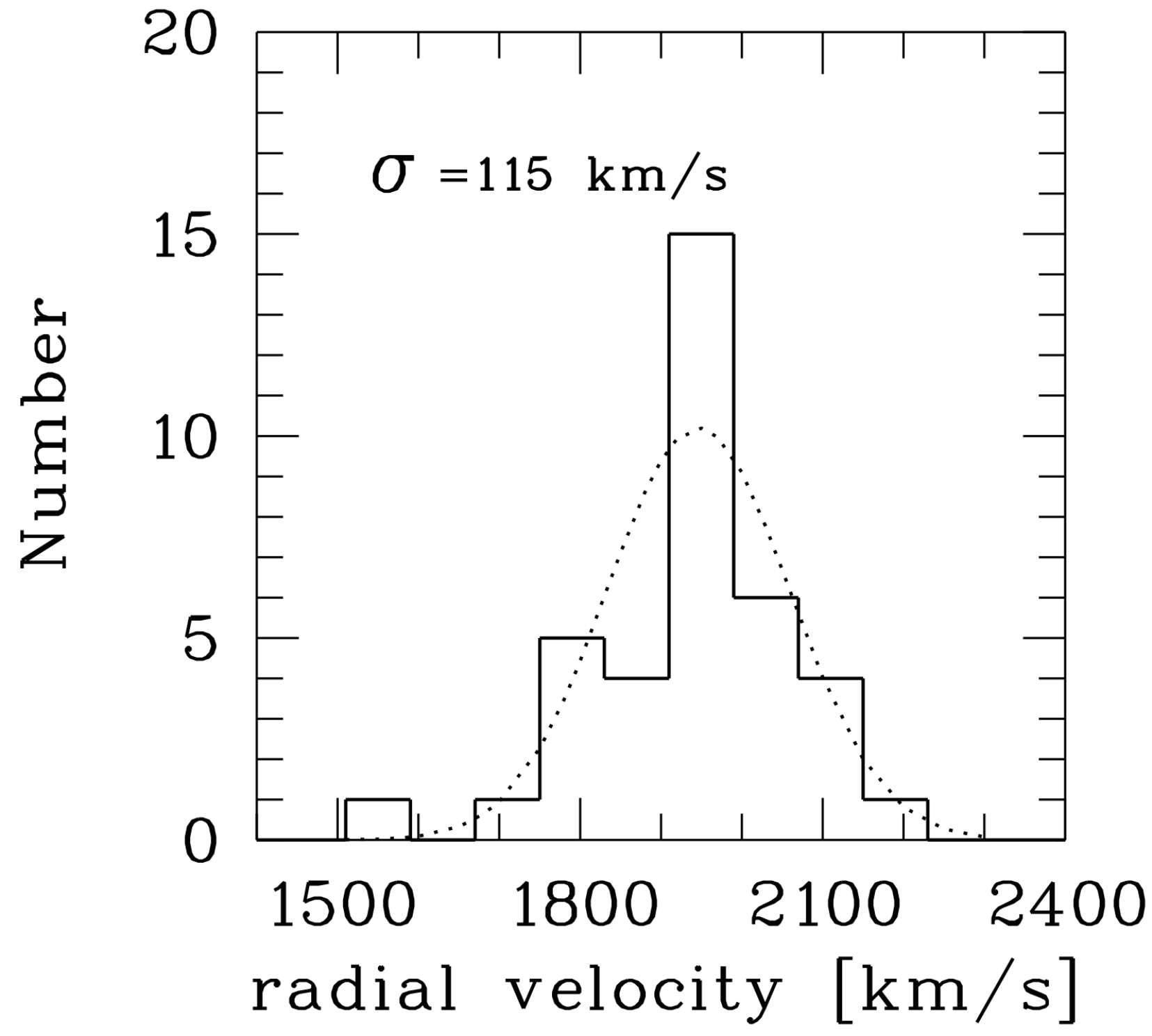}
\caption{The velocity distribution of globular clusters. The dashed line represents the fit of a Gaussian with a dispersion of 115 km/s.  }
\label{fig:gauss}
\end{center}
\end{figure}

A Gauss fit with a fixed central value
of $v_r$ = 1925, which gives less weight   to outliers, results in $\sigma$= 115$\pm$28 km/s, where $\sigma$ is the dispersion of the Gaussian. This Gaussian is shown as the dashed line in
Fig.\ref{fig:gauss}. The  central part is not well fitted. A Wilkinson-Shapiro test for normality gives a p-value of 0.09 and thus does not provide a convincing argument against a normal distribution.
However, the kurtosis of the velocity sample is 4.30, characterising  a distribution that is more peaked than a Gaussian which physically could indicate a radial bias.   More precise statements
are not possible due to the small number of GCs.

A first mass estimation can be tried by inserting the dispersion into the virial theorem (e.g. \citealt{binney08}, equation 4-80)

\begin{equation}
m_{virial} = 2.5 \times r_{half} \times \sigma^2/G
\end{equation}
where $r_{half}$ is the half-mass radius and G=0.0043 (constant of gravitation in units of solar mass, km/s, and parsec). The half-mass radius is not well determined. Adopting 50 kpc as the
radius of the galaxy, the half-mass radius, according to our model (see section \ref{sec:photometry}), is 8.3 kpc, $\sigma$=115 km/s and the corresponding mass is $6.4\times10^{10} M_\odot$. 
The resulting mass-to-light ratio is $M/L_R = 1.05$, much too low for an old/intermediate-age population. A more detailed dynamical analysis arrives at
more reasonable values (see section \ref{sec:jeans}).

%\section{Population properties and velocities}

\subsection{Colour-magnitude diagram and colour distribution}
\label{sec:cmd}
\begin{figure*}[]
\begin{center}
\includegraphics[width=0.8\textwidth]{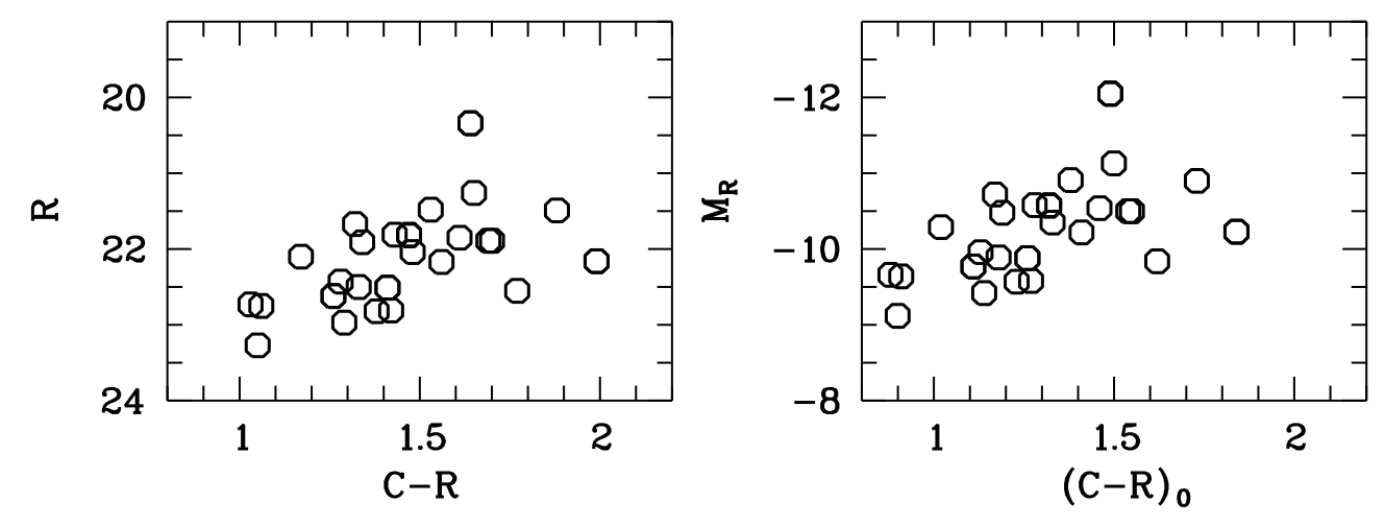}
\caption{Left panel: The colour-magnitude diagram of globular clusters with measured velocities. Right panel: Extinction and reddening corrected and with absolute magnitudes.
}
\label{fig:CMD}
\end{center}
\end{figure*}

Fig.\ref{fig:CMD} shows the CMD of our GC sample with available photometry.  The left panel displays the observed colours and magnitudes, for the dereddened colours and absolute magnitude   of the right panel,
the following numbers apply. The foreground reddening, according to the NED, is E(B-V) = 0.076. The absorption in R is 0.19 mag and the reddening in C-R is 1.97$\times$E(B-V)=0.15 mag \citep{harris77}. These numbers
and the distance modulus m-M=32.23 are used for the calculation of the absolute magnitudes in the right panel of Fig.\ref{fig:CMD}. Given
a  specific frequency of $S_N = 1.2$ \citep{lane13}, which is very low for an elliptical galaxy, it may look surprising to  find so many bright GCs. Particularly striking is the bright object with $M_R$ =  -12.2 mag
(see caption Fig.\ref{fig:slits}.
As will be shown later (see section \ref{sec:on}) through the analysis of the stellar M/L-ratio, NGC 5812 is dominated by light from intermediate-age populations. Many GCs are therefore expected to have higher luminosities than their old counterparts of the same mass.
Comparison with the colour distribution of \citep{lane13} shows that our GCs sample the full colour interval. A further hint to intermediate-age clusters is the fact that there appear clusters with C-R$\approx$0.9
which is too blue for old clusters even with very low metallicites. 

%\subsection{Sample completeness}

\begin{figure}[]
\begin{center}
\includegraphics[width=0.4\textwidth]{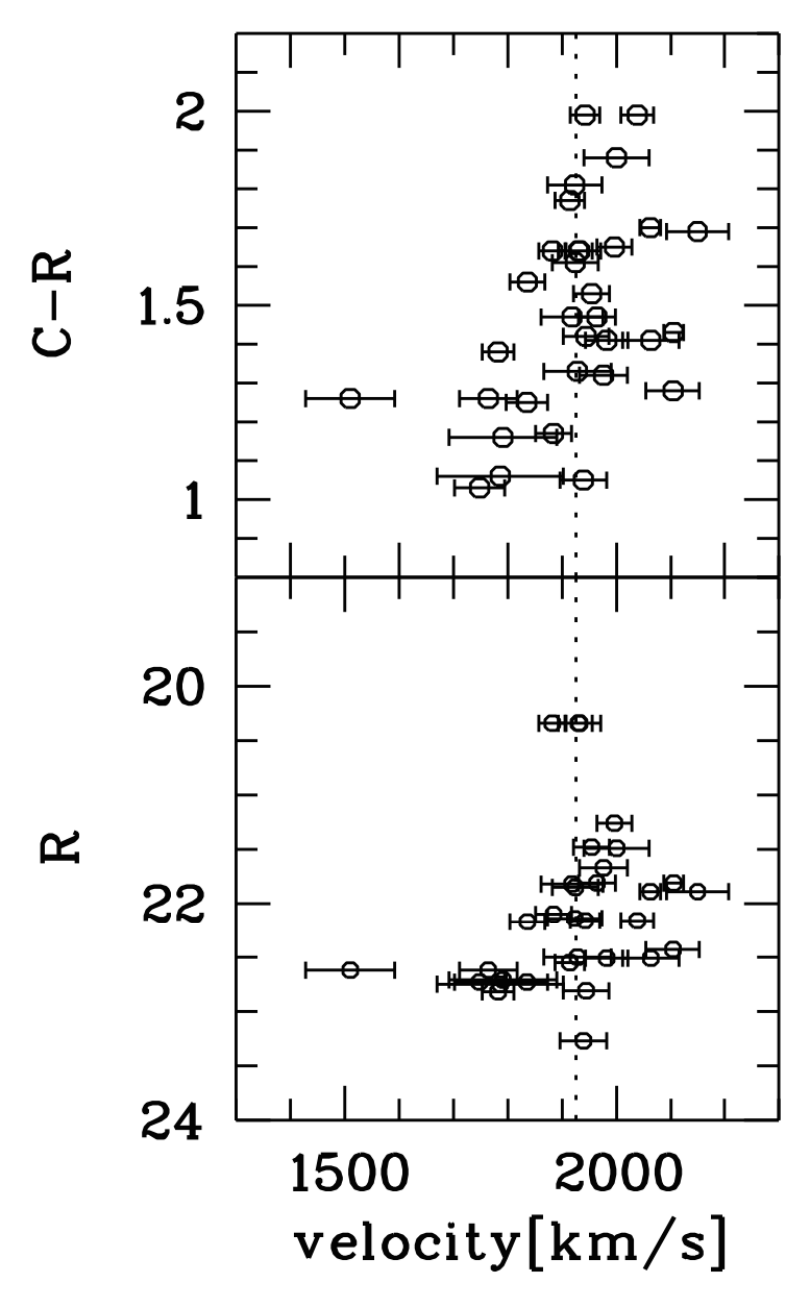}
\caption{Upper panel: Colour C-R vs. radial velocity for all slits. Lower panel: R-magnitudes vs. radial velocities for all slits. The vertical dotted line is the systemic velocity
of NGC 5812.
  }
\label{fig:colvel}
\end{center}
\end{figure}

In the upper panel of Fig.\ref{fig:colvel}, C-R colours  are plotted versus radial velocities.

\subsection{A UCD in an isolated elliptical}
\label{sec:UCD}
There is one strikingly bright object among our GC sample (see caption of Fig.\ref{fig:CMD}).  Unfortunately, the designation Ultracompact Dwarf has become common for GCs brighter than $M_V = -11$  or with $R_{eff}$
larger than 10 pc, although there is no evidence that these objects share galaxy properties (see the introduction in \citealt{fahrion19}).
We combined one spectrum from mask 1 and one from mask 2. The S/N of this combined spectrum (see the appendix for the spectrum) is sufficiently high to permit an analysis. 
We use  the python version of the Penalized Pixel-Fitting (pPXF) method to extract the kinematics and population parameters (if possible) for our spectra. pPXF  was first described by \citet{cappellari04}
and has been upgraded by \citet{cappellari17}, to which the reader is referred for details of the fitting procedure. pPXF uses the MILES stellar population library \citep{vazdekis10}.

pPXF returns
a mass -weighted age of 12 Gyr and a mass-weighted metallicity of [M/H]=-0.5 .  With these values, one would expect a colour of C-R=1.72 according to the Padua models \citep{bressan12}, while we measure
a dereddened colour of C-R=1.50. Uncertainties in the analysis, the applied foreground reddening, the photometric calibration, and the stellar models and the photometric transformations let a difference
of 0.2 mag appear not surprising. A safe statement may be that it is an old and moderately metal-poor object. 
Adopting the above values for age and metallicity, the models give $M/L_R = 4.4$ (with a Chabrier IMF) and the mass follows as $1.6\times10^7 M_\odot$.  This object is unfortunately not on any 
HST image. On Fig.2  of \citet{lane13} one observes that this object might be associated to a stream, but independent observations are demanded before any strong conclusion can be drawn.

%\section{Velocity distributions} 

%\subsection{Velocities, colours, magnitudes}

%\subsection{Velocity dispersion}

%\subsection{Two-dimensional distribution}

\subsection{The companion}

The companion of NGC5812 forms a long tidal tail testifying to  its proximity to NGC 5812 \citep{lane13}. It is nucleated and of early-type. A Washington colour C-R=1.33 is typical for
an old, moderately metal-poor population. The nucleus is not distinguishable in colour from the bulk of the stellar population. 
We put a slit of length 5\arcsec on the dwarf galaxy, but S/N turned out to be insufficient for a population study.
The radial velocity is 1665 km/s,  differing by -260 km/s from the systemic velocity.
 The circular velocity at this projected distance 2.5' (20.3) kpc)  is
 between 190 km/s (stellar mass only) and 270 km/s (MONDian expectation)  and therefore we expect the dwarf on its orbit to be near its pericenter.

in Fig.2 of \citet{lane13}, one   notes the distinct loop of the northern tail that indicates a change of the tail's direction. The easiest explanation would be a projection from a part of the orbit before the apocentre
onto the part after the apocentre. This would suggest the northern arm to be the trailing arm and the southern arm to be the leading arm.
%%In Fig.\ref{fig:dwarf}, we show the slit that we put over
%the galaxy's body together with the spectrum extracted from the full slit length. 
%%The radial velocity is 1665 km/s,  different  by -260 km/s from the systemic velocity.
%This velocity identifies in Fig.2 of \citet{lane13} the southern tidal arm as the leading tail and the northern arm as the trailing tail.  
%The circular velocity at this projected distance 2.5' (20.3) kpc)  is
 %between 190 km/s (stellar mass only) and 270 km/s (MONDian expectation)  and therefore we expect the dwarf on its orbit to be near its pericenter. 
 %In this zone the  tidal tails  mark the orbit itself, while in the apocenter regime, the tails are perpendicular to the
% orbit \citep{lokas15}.. 
 %One   notes the distinct loop of the northern tail that indicates a change of the tail's direction. So the northern tail appears as a projection of the entire orbit from the apocenter to the
 %pericenter whereas the visibility of the southern tail is restricted by the light of NGC 5812.
  As \citet{lane13} note, the plurality of tidal features might be the debris of previous orbits
 of the companion.      
 
Our spectra do not show emission lines,  consistent with the possibility that the dwarf never picked up interstellar gas so that young or intermediate-age
populations do not exist.   

%  The stellar H$\alpha$ absorption is at the expected location, but is accompanied by a red-shifted emission that  does not make the impression of a P Cygni profile (see right panel 
 %in Fig.\ref{fig:dwarf}, in which case the emission should be at the systemic velocity of the host galaxy. The line width matches
 %the expect value of 4.2\AA\ and the velocity is 1900 km/s, close to the systemic velocity of NGC5812. 
% The corresponding
 %[NII]-emission (6549\AA\  or 6583\AA\ ) at the same velocity is not clearly visible.  
 %The dwarf may be a counterpart to the companion of another isolated elliptical galaxy, namely NGC 7796. In this galaxy, the ionised gas forms a ring, probably in relation to intermediate-age
 %globular clusters.      

 See the discussion section for more remarks.

%\section{Analysis of the bright galaxy spectra}
%There are not many spectra 

\section{The galaxy spectra}

We put 17 slits on the galaxy itself in mask 1 and mask 2. To enable sky subtraction, we put sky slits at  large galactocentric radii to ensure minimal galaxy contamination. 
Because the spectral range changes with the frame x-coordinate, we made sure that a given galaxy slit finds its sky slit with  the corresponding spectral range. 
 The slit  locations are given in Table A1.
%We use  the python version of the Penalized Pixel-Fitting (pPXF) method to extract the kinematics and population parameters (if possible) for our spectra. pPXF  has been first described by \citet{cappellari04}
%and has been upgraded by \citet{cappellari17}, to which the reader is referred for details of the fitting procedure. pPXF uses the MILES stellar population library \citep{vazdekis10}.

\subsection{The galaxy slits and their kinematical information}
pPXF offers the possibility to fit 4 moments of the shapes of the absorption lines:  the velocity, the velocity dispersion, the skewness, and the kurtosis (see e.g. section 6.2 of \citealt{schuberth10}). The latter two 
higher moments describe the asymmetric and symmetric deviations from Gaussianity and are  expressed by the respective coefficients of Gauss-Hermite polynomials  (e.g. \citealt{vandermarel93}).
   Fitting only the first two moments is more appropriate, given the limited S/N, but we performed
both fits with four moments without using the higher moments for any analysis. The velocity dispersion in case of fitting four moments is significantly smaller by about 20-30 km/s. The question is what is the "correct" dispersion? The velocity distribution 
whose second moments enters the Jeans-equation is a pure Gaussian only in the case of an isothermal density profile in connection with isotropy. 
In Fig.\ref{fig:jeans}, we show both sets of measured velocity dispersions.
%We cannot exclude systematic shifts, when fitting fourparameters instead of two and  thereforeuse only the higher dispersion value for the dynamical analysis. 
The radial velocities (the heliocentric correction is only -0.2 km/s for mask 1)  are given in Table A1.

\subsection{Stellar populations}
\label{sec:on}
Constraining the population parameters age and metallicity allows to constrain the stellar mass-to-light ratio, assuming the shape of the stellar mass function. 
The GC spectra, due to their low signal-to-noise, are not suitable for such  analyses. Only the brightest globular cluster  and some spectra near the galaxy centre have a suitable brightness. 
For those objects, the results of the population synthesis using pPXF are listed in Table A1. 
%pPXF does not give uncertainties. 
A metallicity gradient is apparent with the central spectra being similar to  solar metallicity, spectra at larger radii showing  lower metallicities typically of [Fe/]=-0.3. A linear regression gives $ [M/H] = -0.1*R[kpc] + 0.083 $.
A difference in age cannot be seen.  All spectra indicate a very old age. Because the pPXF-ages  are older than the standard age of the universe (they are not cosmologically calibrated), we simply adopt 12 Gyr.  Fig.8 of \citet{lane13} shows that the
reality of this metallicity gradient is supported by the colour gradient which  exists within 50\arcsec (6.8 kpc). 

That means that also the M/L of the stellar population shows a gradient and that models with mass-follows-light may be doubtful even in absence of dark matter.      
%ppxf became a popular tool for analysing composite stellar populations and extract kinematic properties of galaxy spectra. For details of the fitting procedure see \cite{}. 

%Table \ref{tab:ppxf} shows the results 
%of the analyses from the V500 cube for three different spectra with different extraction radii. The population syntheses have been performed in the wavelength interval 3800\AA\ - 7000\AA.  We avoid to
%subdivide the field too much in order not to lose S/N and present  3 spectra for three different radial zones. Our photometric work indicates a colour gradient.  We therefore expect lower metallicities at larger
%radii. Indeed the outermost spectrum   

%  Theoretical stellar $M/L$ values are uncertain both for uncertainties in the stellar mass function and for modelling the luminosity of old stellar systems;  see the literature given at
   The webtool  CMD3.6 (http://stev.oapd.inaf.it/cgi-bin/cmd-3.6) provides theoretical M/L-values in many photometric systems (among other things), offering a rich parametrisation. With default parameters and a 
  Chabrier log-normal stellar mass function, one gets the following M/L-values in the R-band: 
For an age of 12 Gyr and solar metallicity, $M/L_R$ = 5.1, for an age of 12 Gyr and [M/H]=-0.4, $M/L_R$ = 4.2. For these values, we adopted the absolute magnitude of the sun as $M_R$ = 4.43 \citep{willmer18}.

\section{Dynamics of NGC 5812}

\subsection{A revised photometric model}
\label{sec:photometry}
Our photometric model from \citet{lane13}, where we fitted the brightness profile with a double beta-model with exponents -1.7 and -0.9, was thought to be a  very good spherical representation of the light distribution.  Unfortunately, the measurements have a calibration error that results in a model which is too bright in R by 0.45 mag. It has, moreover,
a practical disadvantage in that it does not permit a closed analytical formula of the three-dimensional cumulative luminosity, which is useful in numerical integration of Jeans models.
Therefore we use in this work a corrected  formula, which is as good in describing the profile, but   is   a simple beta-model with -1 as the exponent.  Using the photometric data from  \citet{lane13} (their Table A.2.) the surface brightness then is

\begin{equation}
\label{eq:sbmodel}
\mu(R)=-2.5 \times \log \left [a_1 \left (1+\left ( \frac{R}{R_c} \right )^2 \right ) \right ]^{\beta} 
\end{equation}

%original 1.438e-7
with{\bf $a_1$\,=\,7.19$\times$10$^{-8}$, $R_c$\,=\,8.0\arcsec.} 

\begin{figure}[h]
\begin{center}
\includegraphics[width=0.5\textwidth]{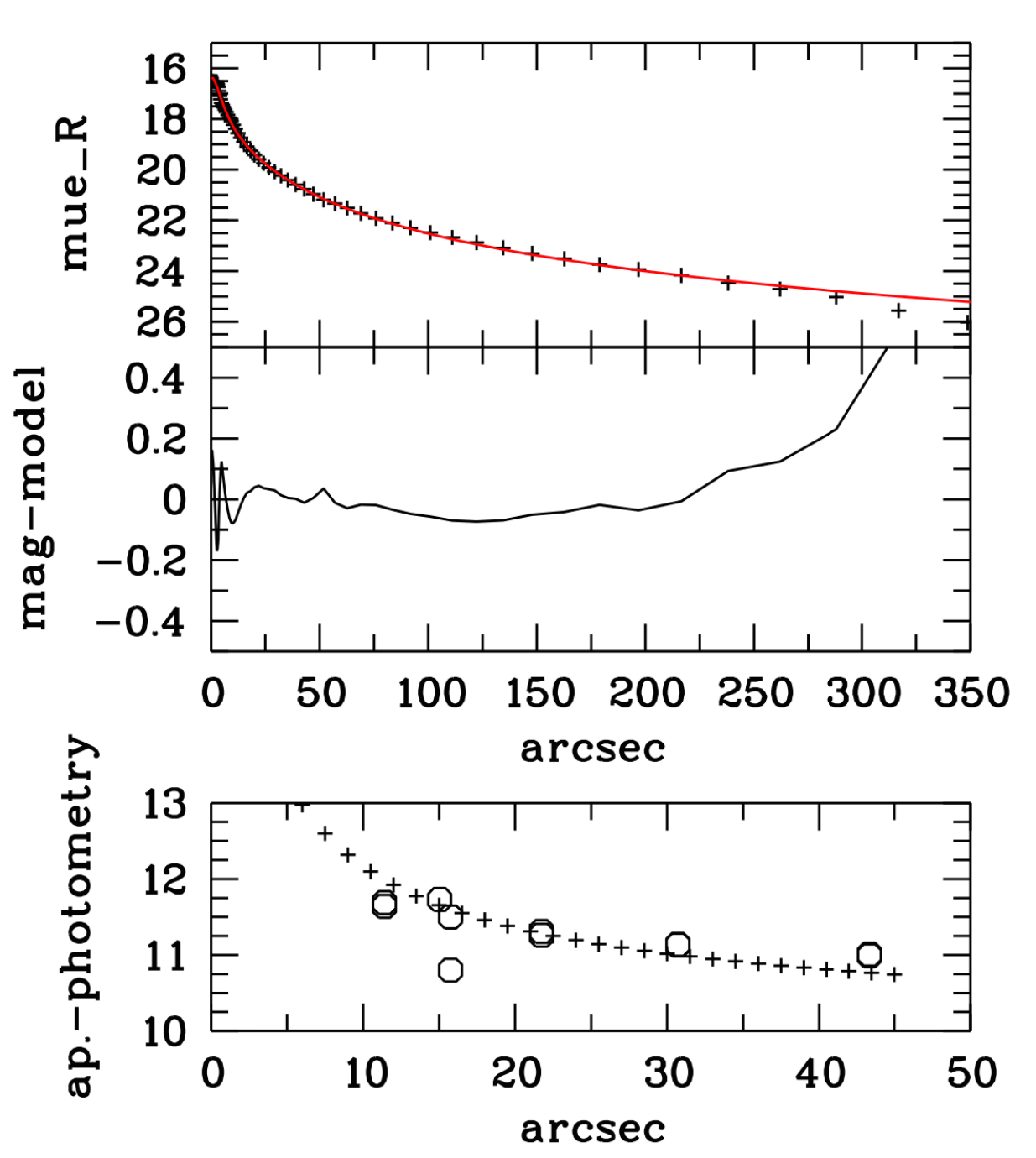}
\caption{ Upper panel: Corrected surface brightness from Paper I and our new formula. Middle panel: Difference between measurements and our model. Lower panel:
numerical aperture  photometry of our model (crosses) is compared with the aperture photometry in the R-band of NGC 5812 in the catalogue of \citet{prugniel98}. 
It shows the validity of our actual photometric calibration.  }
\label{fig:photmodels}
\end{center}
\end{figure}

In Fig.\ref{fig:photmodels} we plot the new surface brightness profile that is now calibrated by the aperture photometry in the R-band from the catalogue of \citet{prugniel98}.
Equation \ref{eq:sbmodel} is an excellent representation of the brightness profile out to 230\arcsec (or 31 kpc). Beyond, the model becomes a bit brighter than the
galaxy, but this happens at a radius where shells already are visible and  sphericity is obviously not longer a good approximation. The lower panel
demonstrates that the model reproduces very well the aperture photometry. 

This model (sometimes called the Modified Hubble Law) can be analytically deprojected to arrive at  equation \ref{eq:depro}.

The  deprojection  gives for the   luminosity density  (unit $L_\odot/pc^3$) 
\begin{equation}
j_0 = f\,  \frac{a_1}{2 r_c} \left [1+ \left ( \frac{r}{r_c} \right )^2 \right ]^{-\frac{3}{2}} 
\label{eq:depro}
\end{equation}
where the radii are now in pc and $f=2.505\times10^{10}$ is the factor that transforms surface brightness into solar luminosities per square pc.
In addition, we apply  a factor 1.19 to correct for foreground absorption in the R-band (see section \ref{sec:cmd}). 
After spherical integration, one obtains for the cumulative mass

\begin{equation}
\begin{split}
M(r) = 4 \pi \,(M/L) \,f  \, \left ( a_1/ \left ( 2 r_c  \right ) \right ) \\ \times \left [ -r  r_c^2/ \sqrt{1+(r/r_c)^2}   + r_c^3 \ln \left ( r/r_c + \sqrt{1+(r/r_c)^2} \right ) \right ]
\end{split}
\label{eq:cumlum}
\end{equation}

If we define 50 kpc to be a cut-off radius (large enough this specific choice  does not affect much the half-light radius), the stellar half-mass radius is 8.5 kpc, which contains a mass of {\bf  $9.97\times10^{10} M_\odot$} adopting 
$M/L_R = 4.2$ to be consistent with an age of 12 Gyr and a metallicity of [M/H]=-0.4. 
 At this radius the MONDian
 mass expectation would be {\bf $1.43\times10^{11} M_\odot$} (equation \ref{eq:mond}), corresponding to a dark matter fraction of 0.27, which can be compared with cosmological simulations.

\subsection{Spherical Jeans models}
\label{sec:jeans}

\begin{figure*}[ht]
\begin{center}
\includegraphics[width=0.8\textwidth]{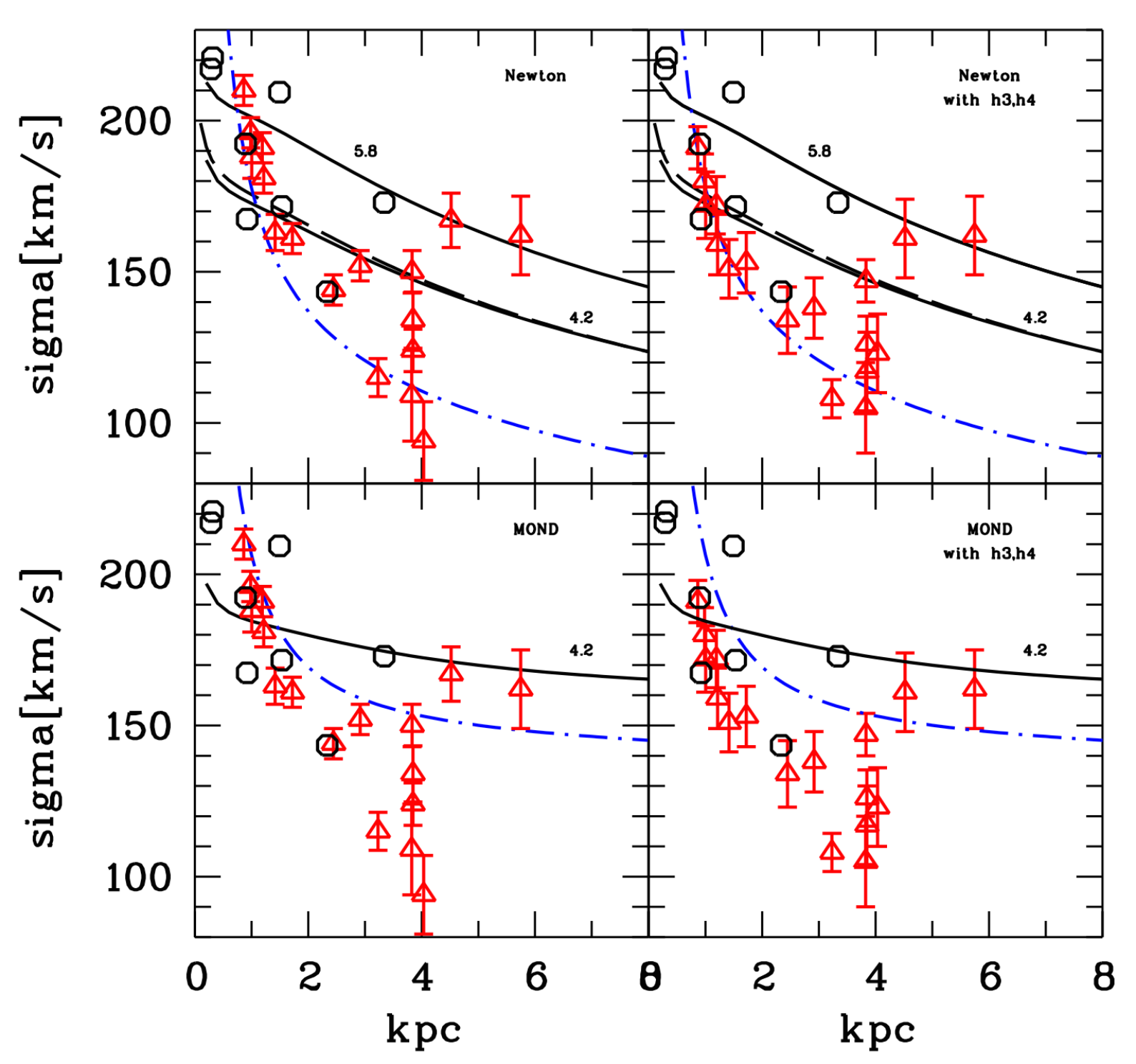}
\caption{ Velocity dispersions of the galaxy light and various spherical Jeans-models of which none is  satisfactory. The circles mark long-slit measurements from \citet{bertin94}. The  triangles are our measurements. {\bf Upper panels:} The upper panels show baryonic Newtonian models. The left panel shows pPXF results, fitting two moments, the right panel fitting four moments. The solid lines are isotropic models with stellar M/L$_R$-values of 5.8 (upper line) and 4.2 (lower line). They do not reproduce the measurements well.
%The dashed line is an isotropic MONDian model with M/L$_R$ = 4.2.
The dashed line (in both panels)   is a model with M/L$_R$=4.2 and 
the  radial anisotropy of \citet{hansen06} (see text) and may represent  the inner dispersion profile, but fails at larger radii. The blue dashed-dotted line (in both upper panels) is a fully radial model, which
seems to fit  most of the measurements, but is unphysical.
Lower panels: The lower panels show the corresponding MONDian models for an M/L$_R$-value of 4.2 (there is no point for showing 5.8).
 They do not fit. See text for more remarks.}
%The blue dot-dashed lines are models (the upper line being the MONDian counterpart of the lower) of two components (see details in the text) and serve as an example of fine-tuning.
%By fine-tuning M/L, anisotropy and MONDian/not-MONDian, one may find a more or less good representation for the inner parts. The wide scatter of velocity dispersion at a given radius rather
%suggests that a purely spherical Jeans model is not an adequate description.  At any case, there is no evidence for a significant spherical halo of dark matter beyond the MONDian expectation.
%In an axisymmetric model, the MOND force may exist perpendicular to the line-of-sight. }
%There is good agreement in the very inner parts, but starting at about 1 kpc, the radial dependence of the dispersion values becomes  widely scattered, which we interpret as real. 
%The solid line is a spherical Jeans-model with M/L=4, based on our revised galaxy light profile,  and without dark matter. It appears not to be a good representation. The lower dashed line  is a model 
%with the radial anisotropy of \citet{hansen06} without dark matter and seems to be a good representation of the inner dispersion profile.  The upper dashed line is a MONDian model (see Section \ref{sec:BTF}).
%Right panel: mass profiles. The open squares mark the mass profile derived from the "tracer mass estimator" assuming isotropy. The open circles denote the mass profile assuming radial anisotropy.
%The dashed lines are the corresponding mass profiles of the left panel.  
\label{fig:jeans}
\end{center}
\end{figure*}

The results from measuring the galaxy spectra can be seen in Fig.\ref{fig:jeans} that shows  the observed velocities and velocity dispersions together with dynamical models.
In the following, we explain in detail what can be seen in Fig.\ref{fig:jeans}. 

The circles are measurements from \citet{bertin94}, which  are
longslit data that in our figure have been folded into one radial coordinate.  The scatter already at small radii is strikingly high. The red triangles are our measurements, the
left panels display the values which resulted from the pPXF-fits with two moments, the right panels from the fits with four moments. There is a small systematic in that the 4-moment fits give
slightly lower dispersions.  We do not think that the values of the higher moments are interpretable
and  do not discuss them further. The main characteristics are the high central dispersion, the rapid decline and the large scatter at larger radii which is in line with the Bertin et al.-values.

The upper panels show baryonic Newtonian models, the lower panels their MONDian correspondence. 
Our models  uses the non-rotating spherical Jean-equation, as we did in previous papers. The Jeans-formalism was presented in many contributions, we refer the reader to \citet{mamon05} and  \citet{schuberth10,schuberth12}. 
 We assign an M/L-ratio (R-band) to   Eq.\ref{eq:cumlum} and calculate the projected velocity dispersions.

We first consider cases of  isotropic models,  which are shown in Fig.\ref{fig:jeans} by  the solid lines.  The upper solid line represents an  M/L$_R$-value of 5.8, motivated partly
by the central population properties and partly by the wish to best reproduce the inner velocity dispersion values. The small central upturn is  caused by a central black hole of $10^9 M_\odot$
that we inserted to illustrate its negligible influence. 
The lower solid line represents $M/L_R$=4.2, motivated by the population gradient. 
It is obvious that these isotropic models are no good descriptions. 

%The MONDian version with $M/L_R$=5.8 would be far off from the observations, so we do not give it.

 A suitable population mix would perhaps look more promising, but in
the framework of spherical models, a better choice would be 
%  A better description is found  lowering the global M/L-value 
  introducing a radial anisotropy.  Pragmatically, we follow \citet{hansen06} who found in  merger simulations 
that the anisotropy is  related to the logarithmic slope of the three-dimensional stellar mass 
distribution by
\begin{equation}
   \beta = 1 -1.15 (1+slope(r)/6).
 \end{equation}  

From our photometric model,  $\beta$ would reach a constant radial anisotropy of +0.4 at about 5 kpc.  A  good approximation for this relation
is the anisotropy profile considered by \citet{mamon05}: $\beta = 0.5(r/(r+r_a))$, $r_a$ being a scale radius with some low value, we choose arbitrarily $r_a$ = 450 pc . Adopting this kind of anisotropy, we
can conveniently apply the formalism given by \citet{mamon05}. 
%For the moment we live with the inconsistency that the luminosity density replaces the total mass density.

% However, merger simulations rather indicate modest radial anisotropies. We use the findings of \citet{hansen06} that the resulting anisotropy of stars in their merger simulations is related to the logarithmic slope of the three-dimensional stellar mass 
%distribution by
%\begin{equation}
%   \beta = 1 -1.15 (1+slope(r)/6).
% \end{equation}  
%     For our photometric model,  $\beta$ reaches a constant radial anisotropy of +0.4 at about 5 kpc.  A  good approximation for this relation
%is the anisotropy profile considered by \citet{mamon05}: $\beta = 0.5(r/(r+r_a))$, $r_a$ being a scale radius with some low value.   
This long-dashed lines in Fig.\ref{fig:jeans}
are calculated with this anisotropy and a global $M/L_R$-value of 4.2. 

This model reproduces the initial decline of the velocity dispersion out to 3 kpc and indicates that within the assumption of sphericity, a radial anisotropy is needed.
 To show the extreme, the blue dashed-dotted line is a fully radial model (kernel ''radial'' of \citealt{mamon05}). This is of course unphysical. Moreover, the central velocity dispersion reaches about 2000  km/s,
but the shape at larger radii is well reproduced. That such unphysical model seems to fit, may be an indication that spherical models are not adequate, but axisymmetric models would need
the knowledge of the full velocity field and are beyond
our scope.

The lower panels show the MONDian versions of the above models.
We calculate the  MONDian circular velocity with  
\begin{equation}
 v_{M} = \sqrt{v_N^2(r)/2 + \sqrt{v_N^4(r)/4 + v_N^2(r)  a_0 r}},
 \label{eq:mond}
\end{equation}
where we adopt $a_0 = 1.35\times10^{-8} cm/sec^2$ \citep{famaey07}  and $v_N$ is the Newtonian circular velocity.  

The MONDian version with $M/L_R$=5.8 would be far off from the observations, so we do not give it. It is also obvious that this kind of MOND models is not satisfactory.
By fine-tuning M/L, anisotropy and MONDian/not-MONDian, one may find a more or less good representation for the inner parts. The wide scatter of velocity dispersion at a given radius rather
suggests that a purely spherical Jeans model is not an adequate description.  At any case, there is no evidence for a significant spherical halo of dark matter.
In an axisymmetric model, the MOND force may exist perpendicular to the line-of-sight.

% The scatter already at small radii is strikingly high.  That also our data show a similar scatter is a strong indication that this scatter is real. 
%Regarding the disturbed morphology of NGC5812, this finding is  not surprising.  
However, a dark matter halo  is not visible under sphericity. The question, whether spherical Jeans-models are adequate at all, is difficult
to answer. In an ideal spherical "Jeans-world" there is a unique relation between velocity dispersion and radius. A significant scatter could indicate  that dynamical equilibrium
is not reached and the MONDian force would be hidden.
% Another possibility would be a flattening along the line-of-sight.

\section{Mass estimators}

\begin{figure*}[ht]
\begin{center}
\includegraphics[width=0.7\textwidth]{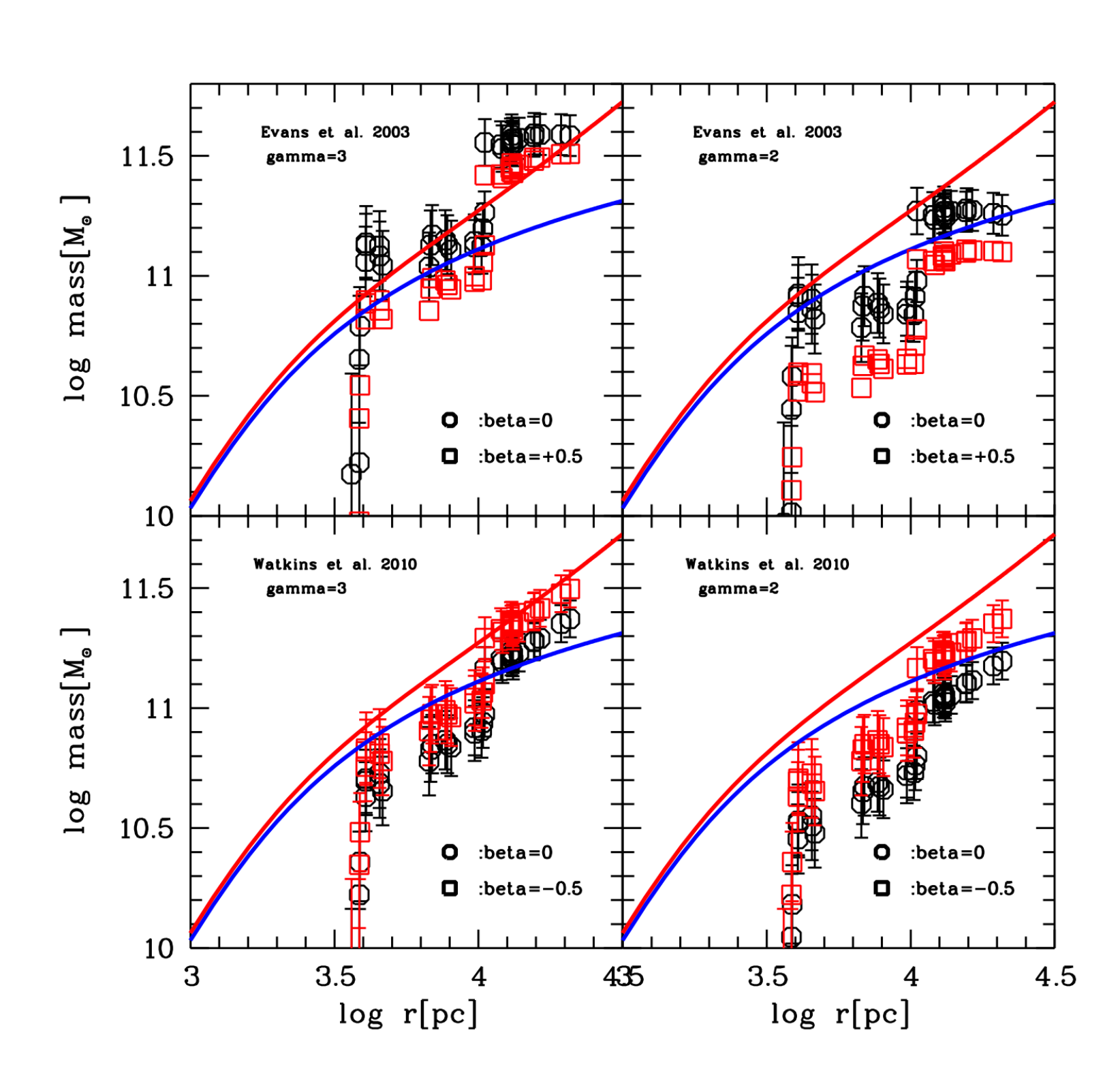}
\caption{Two mass estimators with different parameters, applied consecutively. Sensibly, only the outer points carry sensible mass information. Upper panels:  the estimator of \citet{evans03}. Lower panels: the
estimator of \citet{watkins10}. Left and right panels: the tracer's density power law profile with $\gamma$=3 and $\gamma$=2, respectively. Upper panels: Black symbols stand for isotropy, red for radial anisotropy with
 $\beta$=+0.5. Lower panels: black symbols denote isotropy and red symbols tangential anisotropy with $\beta$= $-$0,5. 
   The black solid lines denote purely baryonic mass profiles for an $M/L_R$ of 4.2 for consistency with Fig.\ref{fig:jeans}. The red solid lines are MOND profiles. Only for a very flat tracer density profile, the data 
are not compatible with MOND. In this case, it would be a dark matter free galaxy.   }
\label{fig:models}
\end{center}
\end{figure*}

For larger radii, we use the GC velocities to further constrain the mass profile. Given the low number of velocities, we do not attempt any Jeans modelling, but apply  ''tracer mass estimators".
These mass estimators have a long history \citep{bahcall81,white81,heisler85}  with the motivation of using only projected radii and radial velocities for a reasonable mass estimation.  
Here we use the estimators proposed by \citet{evans03}
and \citet{watkins10}.  
These mass estimators have the form 

\begin{equation}
\label{eq:massestimator}
\frac{C}{GN}   \times  \sum{ v_{LOS,i}^2 R_i}
\end{equation}
with G the constant of gravitation, N the number of tracers, and  $\rm C = C(\alpha,\beta,\gamma$) being a constant that includes properties of the potential and the tracer population. For further physical reasoning and brevity,
we refer to the original papers.
The parameters are the three-dimensional tracer's density profile described by a power-law exponent $\gamma$ and the orbital anisotropy $\beta$. Furthermore, $\alpha$  is the exponent of
a power-law potential. We assume an isothermal potential, 
which is supported also for elliptical galaxies, so $\alpha=0$.  We calculate the uncertainties of the sum in equation \ref{eq:massestimator} using the individual velocity uncertainties.
However, a major contribution to the  absolute mass uncertainties come from systematics (deviation from virial equilibrium, distance, properties of the tracer populations). Therefore, we show in the four panels of Fig.\ref{fig:models} the relation between the stellar model (blue line) and its MONDian correspondence (red line) and the masses derived with the mass estimators for various parameters. 
$M/L_R$=4.2 has been adopted to be consistent with the Jeans-analysis. 
 
The upper two panels employ the constants given by \citet{evans03}, the lower two panels those by \citet{watkins10}. 
 We apply the mass estimator consecutively,  which  means for the about first 20 velocities  the mass profile is of course not really constrained. We vary $\gamma$, the tracer profile, and 
 the anisotropy parameter. The left panel has $\gamma = 3$, which matches the galaxy light, and the right panel $\gamma=2$.    $\gamma=3$ may describe the inner GC population well whose
distribution should be similar to the galaxy light.   Values a bit less and around $\gamma=2$ are typical for the outer and more metal-poor GCs which are donated through
 the infall of dwarf galaxies, which in our case  of a isolated elliptical will not be many and is a reason for the paucity of the cluster system .    $\gamma$  is not well constrained (see Fig.9 of \citealt{lane13}), but the above values embrace the systematic uncertainty.
 
 In all panels, the black symbols represent isotropic models, and the red symbols represent anisotropic models. 
 
 The red symbols in the upper panel represent radial anisotropies of $\beta=+0.5$, the red symbols in the lower panels  
  represent  tangential anisotropy  of $\beta$=-0.5.

The most general statement is that the GCs do not indicate a significant dark matter halo beyond the MOND expectation. MOND is supported with a  steep number density profile, rather independent from
the anisotropy. A radial anisotropy together with a flatter profile would also be consistent with MOND.   The non-Gaussian velocity distribution speaks for a moderate radial anisotropy, which reflects the anisotropy of the stellar population. A flat profile together with a tangential anisotropy would indicate a galaxy free of dark matter.

%MOND would be stronger supported with a moderate tangential anisotropy (see also Section \ref{sec:bayesian}). A stronger support for MOND must come from a larger sample of dynamical tracers, but  a 
%high dark matter content is discouraged anyway.

An interesting observation is the jump in the mass profile at 10 kpc. One could classify this as a statistical feature without much meaning, but parallel one notes that beyond 10 kpc, only blue,
metal-poor GCs appear which suggests a flatter density profile. That means that until 10 kpc, $\gamma$ would be close to 3 and beyond, $\gamma$ would be closer to 2. That would be consistent with the stellar model
at 10 kpc and would at least soften the jump. 
  
% \section{A tiny central dusty disk}
    
\section{Discussion}

%Wenn das Lichtprofile als Hinweis auf MOND verstanden wird, darf aber M/L nicht wesentlich anders sein, sonst verliert das Argument die Kraft

%\subsectiliton{The GC colour distribution within the globular cluster system}
Our aim with the following discussion is to place our results within the context of the current  literature wisdom. 
%It will turn out that there is no place for them.
\subsection{Comparison with simulated galaxies}
Galaxies like NGC5812 or NGC7507 are so far not identified in cosmological simulations. Older work of \citet{niemi10} does not find a large difference regarding the dark matter content between isolated and non-isolated
elliptical galaxies. The recent project Illustris-TNG can perhaps be regarded as the most complete simulations available. Fig.6 of \citet{lovell18} shall be our reference. Galaxies with  stellar masses like NGC 5812 show a median   
dark matter fraction of 0.7 within one half-mass radius which is not compatible with the present state of data, which permits at most the MONDIan correspondence of 0.27.  One might object that NGC5812 is an elliptical galaxy still in the making and may be better compared with galaxies of higher redshift, but that
is beyond the scope of our paper.

\subsection{Dark matter characteristics and the Baryonic Tully Fisher relation}
\label{sec:BTFR}

%It is of course beyond the scope of our paper to discuss in detail every galaxy that appears deviating from the BTFR. On the long term this will be necessary, but 
%here we restrict ourselves to a few remarks that show the present masses still need confirmation. 
\begin{figure*}[ht]
\begin{center}
\includegraphics[width=0.8\textwidth]{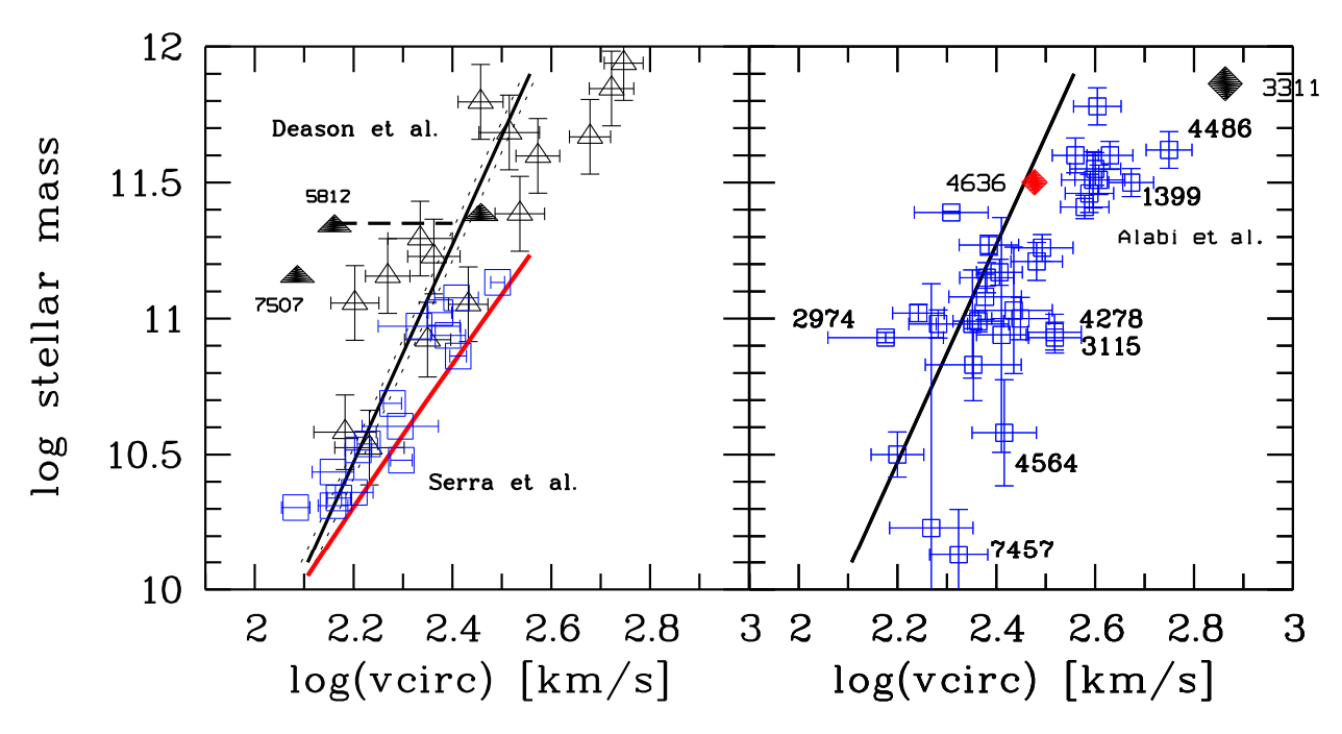}
\caption{This figure demonstrates in a compact manner, what it currently known about masses of early-type galaxies at large radii.  It  displays the total baryonic masses  within
five effective radii vs. the circular velocity at this radius for various samples. The two panels are identical for better readability. The solid lines in both  panels indicate the BTRF
for disk galaxies from \citet{mcgaugh05}, including the 1-$\sigma$-scatter as the dashed lines. Left panel: the triangles denote the sample of  \citet{deason12}, the squares the
sample of \citet{serra16}. The red solid line is the BTFR that results from cosmological simulations \citep{sales17}. The uncertainties have been calculated by error propagation
using the literature information. The intrinsic scatter of the BTFR may be much smaller than the scatter appearing in the graph. Two isolated elliptical galaxies appear with their 
NGC numbers without uncertainties, because there may be unknown systematics (this paper and  \citealt{lane13}). The bar right of NGC 5812 indicates the range which is compatible with
the data. Right panel: blue squares denote the SLUGGs survey \citep{alabi17}. The most
deviant galaxies are indicated by their NGC numbers.  The red diamond is NGC 4636 to which we assign particular weight for its large number of tracers \citep{schuberth12}. The bright deviant galaxies are central cluster galaxies. An additional central cluster galaxy, NGC 3311, is indicated \citep{hilker18}. 
 }
\label{fig:BTFR}
\end{center}
\end{figure*} 

It is now interesting to put these results into the general context of dark matter in early-type galaxies. 
What  is the present understanding  of the relation between dark and luminous matter?
Much work during the last  decades has shown that there is a close relation between the baryonic mass and the dark mass of disk galaxies.  

The well-defined ''baryonic Tully-Fisher relation'' (BTFR) for disk galaxies has been shown various times \citep{mcgaugh05, trachternach09,stark09,mcgaugh12}. 
 %is shown in the left panel in Fig.\ref{fig:BTFR}. This 
The literature  figures should be updated 
%graph in the literature \citep{trachternach09}
 by including six dwarf galaxies from \citet{kurapati18}.   
These dwarfs are found in the Lynx-Cancer void, a region of   very low galaxy density and (within the world of
LCDM) also of low dark matter density. It is  remarkable how well these galaxies fit to the relation given by \citet{mcgaugh05}. It is hard to imagine that such a tight relation for disk galaxies of such different mass
in such different environments shall be the result of galaxy formation in dark matter halos.   

The dynamical properties of disk galaxies  appear {\it as if} the Newtonian law of gravity would be modified according to Milgrom's  prescription \citep{milgrom83a,milgrom83b,bekenstein84}.
Understanding the parent theory of MOND  as a new law of gravitation or as modified inertia, of which MOND is the non-relativistic limit, has inspired
a wealth of theoretical effort (for a review see \citealt{famaey12}; see also sections 1.5.8 and 1.5.9 in \citealt{amendola16}) . 

Interesting thoughts developed  recently within  the framework of a thermodynamical interpretation of gravity. Starting with \citet{bekenstein73} and \citet{hawking75}, who introduced entropy and
temperature in black hole physics, \citet{gibbons77} generalized these concepts for cosmological event horizons.   \citet{jacobson95} showed how the Einstein equation can be perceived
as a thermodynamical equation of state. \citet{strominger96} demonstrated the microscopic origin of the Bekenstein  black hole entropy.
Space-time and gravity may emerge from a microscopic degrees of freedom, e.g. \citet{padmanabhan15,padmanabhan18}. \citet{verlinde11} proposed to describe gravity as an entropic force and \citet{verlinde17}
showed how the MONDian phenomenology may be derived from  this conception in a de Sitter-Universe. \citet{hossenfelder17} provided a covariant version of Verlinde's concept  and \citet{hossenfelder18}
derived a spherically symmetric  non-relativistic limit, which, although physically different, is  the MOND formula in practice. 

 However, theoretical ideas around MOND are evolving, see the recent work of  \citet{milgrom23} and \citet{skordis21}.   
%It is intriguing that these very theoretical considerations result in something observable. However, the concept of gravity as an entropic force is controversial.

If these new physical concepts are valid, every galaxy {\it must} show a MONDian appearance. In particular, 
%One of the manifestations of MOND is the existence of a tight baryonic Tully-Fisher relation (BTFR) among disk galaxies \citep{mcgaugh05,mcgaugh12}.  The left panel in Fig.\ref{fig:BTFR} updates an earlier
%graph in the literature \citep{trachternach09} by including six dwarf galaxies from \citet{kurapati18}.   These dwarfs are found in the Lynx-Cancer void, a region of   very low galaxy density and (within the world of
%LCDM) also of low dark matter density. It is very remarkable how well these galaxies fit to the relation given by \citet{mcgaugh05}.  
%A pestering
%question for the universality of MOND is whether 
elliptical galaxies must fall on the same BTFR as disk galaxies. Apparently, this was not the case in the samples of \citet{gerhard01}
and \citet{magorrian01}, but these studies did not reach out to large radii, where dark matter or mass discrepancies leave their imprints. On the other hand, some elliptical galaxies
have been found, which are well consistent with MOND and therefore fall  onto the BTFR for spiral galaxies \citep{weijmans08, schuberth12, milgrom12}.
%Within the concept of LCDM,  halos of spirals and ellipticals exhibit different structures: at a given total stellar mass, the central densities
%of the dark halos of elliptical are denser than those of spirals.     The early work of \citep{gerhard01} indicated factors up to 25, more modern work find more moderate values
%of 2-5 \citep{napolitano10}.  However, a different halo structure of ellipticals with respect to spirals would make a common BTFR implausible. 
%This relation was
% known only rudimentary when  M. Milgrom gave his first formulation of MOND (\citep{milgrom83a,milgrom83b). The predictive power that MOND has shown over the years (Famaey ..) 
%is partly, but not adequately  acknowledged by the astronomical community \citep{famaey). See also the special issue of the Canadian Journal of Physics.

There is a reason for the fact that the MONDIan phenomenology  among ellipticals is not so clear-cut as for spirals. 
 Spiral galaxies permit a  relatively
easy access to their dynamics by   offering the convenience of a disk with more or less circular orbits.  This disk symmetry is generally missing in ellipticals 
and consequently, mass determinations of elliptical
galaxies can be rather uncertain, even when the kinematical data are good or even excellent. Among the dynamically best known ellipticals  are NGC 3379
and NGC 4494.
In spite of the detailed dynamical knowledge, theoretical models which are consistent with the data span a large variety of dark matter properties 
\citep{morganti13,delorenzi09}, including a MONDian behaviour.

Therefore it is interesting to overcome the individual uncertainty by considering a larger sample of elliptical galaxies  with kinematic data reaching out to
large radii which has been treated homogeneously.    There are now several samples of interest.  These appear in Fig.\ref{fig:BTFR} in the left and the right panel.
The abscissae are the log of the circular velocity at 5 effective radii, the  ordinates are the total baryonic mass (stars+gas) as indicated in the samples. 
%The left panel displays for spiral galaxies updates what is already available in the literature \citep{mcgaugh05,trachternach09}.
%In addition, nine dwarf galaxies with
%HI-disks  from \citet{kurapati18} are plotted with their outermost datapoints of their rotation curves. These dwarfs are found in the Lynx-Cancer void, a region of   very low galaxy density and (within the world of
%LCDM) also of low dark matter density. It is  remarkable howd well these galaxies fit to the relation given by \citet{mcgaugh05},
%demonstrating the independence of the BTFR from environment. 
The black solid lines in both panels denote the BTFR for disk galaxies of \citep{mcgaugh05} $M_{baryonic} = 50 \times v_f^4$ where $M_{baryonic}$ is the baryonic (stars+gas) mass of the galaxy and $v_f$ the
rotational velocity on the flat part of the rotation curve. In the left panel, the uncertainty is indicated by the two dotted lines. The red line represents the mean locus of the BTFR that emerges from cosmological
simulations \citep{sales17}. 
%The other two panels of Fig.\ref{fig:BTFR} contain ellipticals. 
The triangles in the left panel denote the galaxy sample of \citet{deason12}  who homogenised the available kinematic data from the literature and derived masses for 15 ellipticals applying mass indicators based on
tracer populations. 
 The baryonic masses have been calculated by using the mass-to-light ratios of \citet{cappellari11}. The uncertainties have been adopted from the sample descriptions. They
 may be even somewhat underestimated, given the fact that the basic distance uncertainty of an individual galaxy is of the order 15\%-20\%.
   The observed scatter may be largely  explained by these uncertainties.
  Then there is the  interesting  sample  of \citet{serra16}  early-type galaxies with HI-disks who measured the rotation curves  in 16  galaxies (blue squares in Fig.\ref{fig:BTFR}, left panel).
Because the inclination of the HI-disk can be determined with high confidence, the calculated masses may count as more precise than those derived from
tracers like GCs and planetary nebulae.   They  point out the  strong coupling between dark and luminous matter  by the relation
\begin{equation}
   V_{circular}(5 R_{e})   = 1.33 \times \sigma_e 
\label{eq:serra}   
\end{equation}
where $R_e$ is the effective radius and $\sigma_e$ the LOV velocity dispersion within one $R_e$. This relation varies the well-known relation $V_{circular} = \sqrt{3} \sigma_{LOS}$
for an isothermal, isotropic  system. It is obvious that they rediscovered MOND for their sample. These two samples are well consistent with the BTFR of disk galaxies, except for our
isolated ellipticals and some
massive galaxies that in fact are prominent central galaxies. 
%The solid line is not a fit, but the BTFR for spirals from \citet{mcgaugh05}.

 Another even larger sample of 34 ellipticals has  been provided by the SLUGGS survey \citep{alabi17} that  uses GC radial velocities for mass determinations at larger radii (blue symbols
 in the right panel of Fig.\ref{fig:BTFR}).
 The total number of GC velocities is impressively 2500 and some galaxies have several hundred GCs to be used as tracers for the potential.
 The mean locus of the SLUGGs-sample is clearly displaced from the spiral BTFR. The most deviating galaxies are indicated and commented on in the next section.
 In many cases there are reasons to doubt the high masses. The red diamond is NGC 4636, whose location is based on the velocities of 460 globular clusters \citep{schuberth12} and
 therefore has particular weight.
% The red symbols are the ATLAS 3D galaxies, where simply equation \ref{eq:serra} has been applied.
 %Only the SLUGGS survey provide serious deviators from a universal BTFR relation.

\subsection{Galaxies deviating from the BTFR}

The indications for a general MONDian phenomenology are so strong that the deviators from the MONDian relation are at least as interesting as  those agreeing. 
It is of course beyond the scope of our paper to discuss in detail every galaxy that appears deviating from the BTFR. On the long term this will be necessary, but 
here we restrict ourselves to a few remarks that suggest that  the presented masses still need confirmation. 

\subsubsection{Dark matter dominated galaxies}
\label{sec:dmdomination}
 If the new concepts of gravity are viable,  there is no reason to
expect dark matter at all, meaning that it is potentially possible to falsify this proposed nature of gravity just by a quick look at Fig.\ref{fig:BTFR}, where there are many examples deviating from the BTFR. 
Therefore the question is important, whether the deviations mean solid and irrefutable facts or whether these galaxy masses are due to incomplete data and/or systematic effects that are not yet understood.  
%clearly  cheap, but obviously we cannot Then we have the new situation in the literature that the deviators from the MONDian relation are at least as interesting as  those agreeing.  iscuss each galaxy individually.  Then we have the new situation in the literature that the deviators from the MONDian relation are at least aot account s interesting as  those agreeing.  
The obviously deviating galaxies in the Deason et al. sample survey are NGC 1399 and M87, both central galaxies of a cluster or subcluster, respectively. It is common wisdom (perhaps erroneous) that MOND does
not account for all dark matter in galaxy clusters (e.g. \citealt{angus08}), and therefore central galaxies are particularly expected to host huge dark matter halos. NGC1399 is a rather mild example \citep{richtler08}. NGC 3311 as
the central galaxy of the HydraI cluster with its strongly rising velocity dispersion is a stronger example of a galaxy that has been putatively associated with a large cored halo \citep{richtler11}.
However, recent work has shown that the rising velocity dispersion may rather be the result of mixing of different tracer populations in a very inhomogeneous velocity field \citep{hilker18}.  It cannot be shown on the
one hand that
NGC 3311 is MONDian, but on the other hand there is no "clean Jeans-world" and no spherical symmetry, so mass determinations based on Jeans-models are doubtful.

% In the sample
%of \citet{gerhard01}, the isolated elliptical NGC 7507 was the most dark matter dominated galaxy. In strong contrast, \citet{salinas12,lane15} modelled NGC  7507 with much better data without any dark matter.
%Another isolated elliptical  with contradicting dark matter content is NGC 4697, whose possible dark matter halo  was labelled "inconspicious"  by \citet{mendez09}, using the kinematics of over 500 PNe. 

%This object appeared as one of the most dark matter
%dominated galaxies in \citet{alabi16} and still containing a lot of dark matter in \citet{alabi17} on the basis of 20 GC velocities.  

Other galaxies with dark matter content much higher than allowed by the BTFR are  NGC 7457, 4564, 3115, 4278. NGC 4278 is also a galaxy in the Serra et al. sample and  there fits well to the BTFR.
%NGC 4564 falls onto the BTFR if we {\it remove} the dark matter, which it is not permitted, because in this case  $M_{baryonic} = 23.2 R_{eff}^2$ contradicting the structure of early-type galaxies.
  NGC 7457 is the faintest   
galaxy in the entire sample. %Similar investigations for other examples like NGC 6166 \citep{bender} are pending.

%How uncertain an individual galaxy mass may be shows the case of NGC4697. \cite

The data quality plays a role and also our present sample certainly leaves room for improvement.

\subsubsection{Dark matter poor galaxies}
%While the samples of \citet{deason12} and \citet{serra15} define a straight BTFR, deviators are found in the SLUGGS sample. l 
%It is of course beyond our scope to discuss every galaxy in detail (which eventually will be necessary), but a few remarks are interesting.

In the sample
of \citet{gerhard01}, the isolated elliptical NGC 7507 was the most dark matter dominated galaxy. In strong contrast, \citet{salinas12,lane15}  using much more complete data, were able to produce 
 models of NGC 7507 that did not require  any dark matter.
Another isolated elliptical  with contradicting dark matter content is NGC 4697, whose possible dark matter halo  was labelled "inconspicious"  by \citet{mendez09}, using the kinematics of over 500 PNe.  This object appeared as one of the most dark matter
dominated galaxies in \citet{alabi16} and still containing a lot of dark matter in \citet{alabi17} on the basis of 20 GC velocities.  
%The data quality plays a role and our present sample certainly leaves room for improvement.
Another dark matter-poor galaxy is NGC 2974, which is also in the Serra et al. sample, where it fits well to the BTFR. See also \citet{weijmans08} who describe this galaxy as being MONDian.

NGC 5812 may be another example where a low number of tracers together with  a system still approaching an equilibrium does not permit to derive a trustworthy mass. Complete velocity fields of the outer
regions of early-type galaxies
can be observed with large field IFUs, and may help to understand better their dynamics.

% examples are the extreme deviating galaxies that  are not  central cluster galaxies. As evident from Fig.\ref{fig??}, these  is NGC 2974 on the left side of the spiral
%relation.   Other work
%rather placed it onto the relation \citep{weijmans} and the circular velocity of \citet{serra16} qualifies it as an exceptionally dark matter rich galaxy. On the right side, we find NGC 7457, 4564, 3115, 4278.  
%that it   

\subsection{X-ray masses of isolated ellipticals}
The data from Fig.8 are of stellar dynamical origin. However, many early-type
galaxies are X-ray bright, and therefore under the assumption of hydrostatical equilibrium,
provide the possibility to derive masses, among them isolated ellipticals. We do  not aim at completeness, but
refer the reader to some particular cases, where X-ray analyses indeed indicate the location
of galaxies on the BTFR.

That the two isolated elliptical galaxies  NGC 720 and NGC 1521 obey the MONDian prediction has been shown by \citet{milgrom12}, (non-MONDian analyses and  X-ray data from \citealt{humphrey11} and \citealt{humphrey12}). 
 The mass modelling of NGC 7796  \citep{richtler15} (X-ray data from \citealt{osullivan07}) showed at least consistency with the MONDian prediction.
 More X-ray masses of early-type
galaxies in the \mbox{MONDian} context has been listed by \citep{lelli17}.  Their
thorough study of the "Radial Acceleration Relation of Galaxies" includes 240
galaxies of various types, among them also 25 early-type galaxies (their Table
3), of which 16
belong to the kinematic sample of \citet{serra16} and entered already Fig.8.
Among the remaining X-ray based objects is NGC 6482 remarkable as a further
 isolated elliptical on the BTFR.
 All these data let  the ''pure baryonic'' galaxy NGC 7507 so far appear
as a not-understood exception.

%\subsection{An easy distance indicator for early-type galaxies}
%We note in passing that $sigma_e$ serves as a good distance indicator
\subsection{Bayesian fits to rotation curves}
\label{sec:bayesian}
The MOND phenomenology appears with a different flavour as a "universal acceleration scale" in disk galaxies \citep{lelli17}. This scale then should also exist in elliptical galaxies. An alternative approach  to 
dynamics of elliptical galaxies has been to fit  "rotation curves",  by means of bayesian techniques, using as input data the light profiles and projected velocity dispersions and fitting  a sequence of parameters with certain 
priors and assumption, including mass-to-light  ratios and the anisotropy. The results so far are controversal. \citet{chae20} found indeed a universality of the acceleration scale.
However, the Bayesian techniques has been critised by \citet{li21} who cautioned the use of degenerated parameters and demonstrated that G, the constant of gravitation, may turn out different from galaxy to
galaxy. 

Recently, \citet{chan22} found for a sample of MaNGA galaxies  the acceleration scale by roughly a factor 1.5 higher than that for spiral galaxies. They boldly conclude that there is no universal acceleration
among galaxies and take that as a falsification  for MOND.  However, they found for most of their round galaxies a significant tangential anisotropy. To our knowledge, such an anisotropy has never been observed elsewhere 
in the literature of the dynamics of elliptical galaxies, so its reality as a characteristic for elliptical galaxies may be doubted. Because this value appears as $-2\times \beta $ in the  Jeans equation for the galaxy mass, it may well account for the difference to the acceleration scales of spirals. Therefore we  suspect that this may be an example for Li and al.' s precaution.

%, the functional  descriin eIn favour of a common 
 
\subsection{Concluding remarks}
These examples show that the present data do not permit a final convincing conclusion  whether or not early-type galaxies fall onto the BTRF of spiral galaxies. This seems to be a sample
property,  some samples agreeing with the MONDian prediction, others not. However, there are indications that data quality and completeness play an important role.   
Central galaxies seem to be clearly not MONDian,  but NGC 3311 may be an example for  the possibility  that tracer properties can fake a huge dark halo that in reality does not exist.
At the isolated ellipticals  end, NGC 7507 stands still as a candidate galaxy without dark matter. Perhaps the observation of a full velocity field will bring clarity.

\section{Summary and conclusions}

We investigated the dynamics of the isolated elliptical NGC5812 with a database of spectroscopic Gemini/GMOS masks, using
the galaxy light and globular clusters as dynamical tracers. 

We could identify 24 globular clusters at galactocentric distances between 10 kpc and
23 kpc  (the adopted distance is 28 Mpc) and applied the tracer mass estimators of \citet{evans03} and \citet{watkins10}.
As for the galaxy light,  we measured velocities and velocity dispersion values out to a distance of 6 kpc. 
 We revised the photometric model from previous works.   NGC 5812  has  a spherical appearance, but  a subtraction of a spherical model leaves  significant residuals, so one may expect the spherical  symmetry to be violated  as 
  previous long slit spectroscopy already suggested. While the  velocity dispersion profile
 seems to be well defined out to 2 kpc, one finds a large scatter at larger radii. Therefore we refrained from stacking spectra at different position angles.
The central spectra show an old, metal-rich population showing a metallicity gradient.
 We constructed several Jeans-model mass profiles, which assume spherical symmetry. We need   a stellar mass-to-light ratio of $M/L_R = 4.0$ in combination with a slight radial bias for reproducing
the central velocity dispersion profile.  Isotropic models do not work well. This might correspond to the velocity distribution of globular clusters that is not purely Gaussian, but exhibits a significant kurtosis.
We then can compare the baryonic mass with the dynamical mass, which ideally should be manifest in the kinematics of the globular clusters. 
Our mass estimators assume spherical symmetry (which also means a  smooth population of phase space) and virial equilibrium. We find consistency with  the MONDian expectation, but purely baryonic
models are not excluded.  However, the data seem to exclude a massive dark halo. 

One of the globular clusters is extraordinarily bright $M_R \approx -12.2$.  We estimate its mass to be $1.6\times10^7 M_\odot$. We also could measure a radial velocity for  the dwarf companion with a tidal tail which indicates a very elongated orbit as
expected.

%A model for the galaxy light without dark matter is in very good agreement with the isotropic globular cluster model, while we expect a dark matter content that obeys the baryonic
%Tully-Fisher relation. 

The case for a MONDian phenomenology also among early-type galaxies  has become so strong (e.g. \citealt{lelli17}) that deviations from a
MONDian behaviour appear more interesting than agreements, because there is an astrophysical problem to solve. In our previous work, NGC 7507  already appeared as an dark matter-free galaxy. 
We embed the results for NGC 5812 in the present literature for dark matter in early-type galaxies,  using the galaxy samples of \citet{deason12} and \citet{serra16}, which strongly argue for the fact
that elliptical
galaxies fall onto the same baryonic Tully-Fisher relation as spiral galaxies, 
 The sample of ellipticals of  \citet{alabi17} (the SLUGGs-survey)  does not show that so clearly. 

 The (unavoidable) deficiency of our analysis is beside the relative small number of probes the application of mass estimators that assume spherical symmetry and dynamical equilibrium.    
 %Remarks on the isolation and that it may be the opposite effect to central galaxies. 
  It  may be that the velocity dispersion field does not contain the full kinematical information and that a full velocity map is necessary to reliably derive a  mass profile together with more insight in its internal structure.

\section*{Acknowledgements}

TR acknowledges financial support   from FONDECYT project Nr. 1100620, and
from the BASAL Centro de Astrofisica y Tecnologias
Afines (CATA) PFB-06/2007. TR thanks ESO for several invitations as science visitor,
where a major part of this work has been accomplished. 
TR thanks Mike Fellhauer for a class on tidal tails.
We are thankful to Bryan Miller for essential help with the GMOS data reductions.

\bibliography{N5812new_clean}

\begin{thebibliography}{}

\bibitem [\protect \citeauthoryear {%
{Alabi}%
\ \protect \BOthers {.}}{%
{Alabi}%
\ \protect \BOthers {.}}{%
{\protect \APACyear {2016}}%
}]{%
alabi16}
\APACinsertmetastar {%
alabi16}%
\begin{APACrefauthors}%
{Alabi}, A\BPBI B.%
, {Forbes}, D\BPBI A.%
, {Romanowsky}, A\BPBI J.%
\ et al.\end{APACrefauthors}%
\unskip\
\newblock
\APACrefYearMonthDay{2016}{{\APACmonth{08}}}{},
\newblock
\unskip
\newblock
\APACjournalVolNumPages{\mnras}{460}{}{3838-3860}.
\newblock
\begin{APACrefDOI} \doi{10.1093/mnras/stw1213} \end{APACrefDOI}
\PrintBackRefs{\CurrentBib}

\bibitem [\protect \citeauthoryear {%
{Alabi}%
\ \protect \BOthers {.}}{%
{Alabi}%
\ \protect \BOthers {.}}{%
{\protect \APACyear {2017}}%
}]{%
alabi17}
\APACinsertmetastar {%
alabi17}%
\begin{APACrefauthors}%
{Alabi}, A\BPBI B.%
, {Forbes}, D\BPBI A.%
, {Romanowsky}, A\BPBI J.%
\ et al.\end{APACrefauthors}%
\unskip\
\newblock
\APACrefYearMonthDay{2017}{{\APACmonth{07}}}{},
\newblock
\unskip
\newblock
\APACjournalVolNumPages{\mnras}{468}{}{3949-3964}.
\newblock
\begin{APACrefDOI} \doi{10.1093/mnras/stx678} \end{APACrefDOI}
\PrintBackRefs{\CurrentBib}

\bibitem [\protect \citeauthoryear {%
{Amendola}%
\ \protect \BOthers {.}}{%
{Amendola}%
\ \protect \BOthers {.}}{%
{\protect \APACyear {2016}}%
}]{%
amendola16}
\APACinsertmetastar {%
amendola16}%
\begin{APACrefauthors}%
{Amendola}, L.%
, {Appleby}, S.%
, {Avgoustidis}, A.%
\ et al.\end{APACrefauthors}%
\unskip\
\newblock
\APACrefYearMonthDay{2016}{{\APACmonth{06}}}{},
\newblock
\unskip
\newblock
\APACjournalVolNumPages{ArXiv e-prints}{}{}{}.
\PrintBackRefs{\CurrentBib}

\bibitem [\protect \citeauthoryear {%
{Angus}%
, {Famaey}%
\BCBL {}\ \BBA {} {Buote}%
}{%
{Angus}%
\ \protect \BOthers {.}}{%
{\protect \APACyear {2008}}%
}]{%
angus08}
\APACinsertmetastar {%
angus08}%
\begin{APACrefauthors}%
{Angus}, G\BPBI W.%
, {Famaey}, B.%
\BCBL {}\ \BBA {} {Buote}, D\BPBI A.%
\end{APACrefauthors}%
\unskip\
\newblock
\APACrefYearMonthDay{2008}{{\APACmonth{07}}}{},
\newblock
\unskip
\newblock
\APACjournalVolNumPages{\mnras}{387}{4}{1470-1480}.
\newblock
\begin{APACrefDOI} \doi{10.1111/j.1365-2966.2008.13353.x} \end{APACrefDOI}
\PrintBackRefs{\CurrentBib}

\bibitem [\protect \citeauthoryear {%
{Bahcall}%
\ \BBA {} {Tremaine}%
}{%
{Bahcall}%
\ \BBA {} {Tremaine}%
}{%
{\protect \APACyear {1981}}%
}]{%
bahcall81}
\APACinsertmetastar {%
bahcall81}%
\begin{APACrefauthors}%
{Bahcall}, J\BPBI N.%
\BCBT {}\ \BBA {} {Tremaine}, S.%
\end{APACrefauthors}%
\unskip\
\newblock
\APACrefYearMonthDay{1981}{{\APACmonth{03}}}{},
\newblock
\unskip
\newblock
\APACjournalVolNumPages{\apj}{244}{}{805-819}.
\newblock
\begin{APACrefDOI} \doi{10.1086/158756} \end{APACrefDOI}
\PrintBackRefs{\CurrentBib}

\bibitem [\protect \citeauthoryear {%
J.~{Bekenstein}%
\ \BBA {} {Milgrom}%
}{%
J.~{Bekenstein}%
\ \BBA {} {Milgrom}%
}{%
{\protect \APACyear {1984}}%
}]{%
bekenstein84}
\APACinsertmetastar {%
bekenstein84}%
\begin{APACrefauthors}%
{Bekenstein}, J.%
\BCBT {}\ \BBA {} {Milgrom}, M.%
\end{APACrefauthors}%
\unskip\
\newblock
\APACrefYearMonthDay{1984}{{\APACmonth{11}}}{},
\newblock
\unskip
\newblock
\APACjournalVolNumPages{\apj}{286}{}{7-14}.
\newblock
\begin{APACrefDOI} \doi{10.1086/162570} \end{APACrefDOI}
\PrintBackRefs{\CurrentBib}

\bibitem [\protect \citeauthoryear {%
J\BPBI D.~{Bekenstein}%
}{%
J\BPBI D.~{Bekenstein}%
}{%
{\protect \APACyear {1973}}%
}]{%
bekenstein73}
\APACinsertmetastar {%
bekenstein73}%
\begin{APACrefauthors}%
{Bekenstein}, J\BPBI D.%
\end{APACrefauthors}%
\unskip\
\newblock
\APACrefYearMonthDay{1973}{{\APACmonth{04}}}{},
\newblock
\unskip
\newblock
\APACjournalVolNumPages{\prd}{7}{}{2333-2346}.
\newblock
\begin{APACrefDOI} \doi{10.1103/PhysRevD.7.2333} \end{APACrefDOI}
\PrintBackRefs{\CurrentBib}

\bibitem [\protect \citeauthoryear {%
{Bertin}%
\ \protect \BOthers {.}}{%
{Bertin}%
\ \protect \BOthers {.}}{%
{\protect \APACyear {1994}}%
}]{%
bertin94}
\APACinsertmetastar {%
bertin94}%
\begin{APACrefauthors}%
{Bertin}, G.%
, {Bertola}, F.%
, {Buson}, L\BPBI M.%
\ et al.\end{APACrefauthors}%
\unskip\
\newblock
\APACrefYearMonthDay{1994}{{\APACmonth{12}}}{},
\newblock
\unskip
\newblock
\APACjournalVolNumPages{\aap}{292}{}{381-391}.
\PrintBackRefs{\CurrentBib}

\bibitem [\protect \citeauthoryear {%
{Bertone}%
}{%
{Bertone}%
}{%
{\protect \APACyear {2010}}%
}]{%
bertone10}
\APACinsertmetastar {%
bertone10}%
\begin{APACrefauthors}%
{Bertone}, G.%
\end{APACrefauthors}%
\unskip\
\newblock
\APACrefYearMonthDay{2010}{{\APACmonth{11}}}{},
\newblock
\unskip
\newblock
\APACjournalVolNumPages{\nat}{468}{}{389-393}.
\newblock
\begin{APACrefDOI} \doi{10.1038/nature09509} \end{APACrefDOI}
\PrintBackRefs{\CurrentBib}

\bibitem [\protect \citeauthoryear {%
{Bertone}%
\ \BBA {} {Hooper}%
}{%
{Bertone}%
\ \BBA {} {Hooper}%
}{%
{\protect \APACyear {2018}}%
}]{%
bertone18b}
\APACinsertmetastar {%
bertone18b}%
\begin{APACrefauthors}%
{Bertone}, G.%
\BCBT {}\ \BBA {} {Hooper}, D.%
\end{APACrefauthors}%
\unskip\
\newblock
\APACrefYearMonthDay{2018}{{\APACmonth{10}}}{},
\newblock
\unskip
\newblock
\APACjournalVolNumPages{Reviews of Modern Physics}{90}{4}{045002}.
\newblock
\begin{APACrefDOI} \doi{10.1103/RevModPhys.90.045002} \end{APACrefDOI}
\PrintBackRefs{\CurrentBib}

\bibitem [\protect \citeauthoryear {%
{Bertone}%
, {Hooper}%
\BCBL {}\ \BBA {} {Silk}%
}{%
{Bertone}%
\ \protect \BOthers {.}}{%
{\protect \APACyear {2005}}%
}]{%
bertone05}
\APACinsertmetastar {%
bertone05}%
\begin{APACrefauthors}%
{Bertone}, G.%
, {Hooper}, D.%
\BCBL {}\ \BBA {} {Silk}, J.%
\end{APACrefauthors}%
\unskip\
\newblock
\APACrefYearMonthDay{2005}{{\APACmonth{01}}}{},
\newblock
\unskip
\newblock
\APACjournalVolNumPages{\physrep}{405}{}{279-390}.
\newblock
\begin{APACrefDOI} \doi{10.1016/j.physrep.2004.08.031} \end{APACrefDOI}
\PrintBackRefs{\CurrentBib}

\bibitem [\protect \citeauthoryear {%
{Bertone}%
\ \BBA {} {Tait}%
}{%
{Bertone}%
\ \BBA {} {Tait}%
}{%
{\protect \APACyear {2018}}%
}]{%
bertone18a}
\APACinsertmetastar {%
bertone18a}%
\begin{APACrefauthors}%
{Bertone}, G.%
\BCBT {}\ \BBA {} {Tait}, T\BPBI M\BPBI P.%
\end{APACrefauthors}%
\unskip\
\newblock
\APACrefYearMonthDay{2018}{{\APACmonth{10}}}{},
\newblock
\unskip
\newblock
\APACjournalVolNumPages{\nat}{562}{7725}{51-56}.
\newblock
\begin{APACrefDOI} \doi{10.1038/s41586-018-0542-z} \end{APACrefDOI}
\PrintBackRefs{\CurrentBib}

\bibitem [\protect \citeauthoryear {%
{Binney}%
\ \BBA {} {Tremaine}%
}{%
{Binney}%
\ \BBA {} {Tremaine}%
}{%
{\protect \APACyear {2008}}%
}]{%
binney08}
\APACinsertmetastar {%
binney08}%
\begin{APACrefauthors}%
{Binney}, J.%
\BCBT {}\ \BBA {} {Tremaine}, S.%
\end{APACrefauthors}%
\unskip\
\newblock
\APACrefYear{2008},
\newblock
\APACrefbtitle {{Galactic Dynamics: Second Edition}} {{Galactic Dynamics:
  Second Edition}}.
\PrintBackRefs{\CurrentBib}

\bibitem [\protect \citeauthoryear {%
{Bressan}%
\ \protect \BOthers {.}}{%
{Bressan}%
\ \protect \BOthers {.}}{%
{\protect \APACyear {2012}}%
}]{%
bressan12}
\APACinsertmetastar {%
bressan12}%
\begin{APACrefauthors}%
{Bressan}, A.%
, {Marigo}, P.%
, {Girardi}, L.%
, {Salasnich}, B.%
, {Dal Cero}, C.%
, {Rubele}, S.%
\BCBL {}\ \BBA {} {Nanni}, A.%
\end{APACrefauthors}%
\unskip\
\newblock
\APACrefYearMonthDay{2012}{{\APACmonth{11}}}{},
\newblock
\unskip
\newblock
\APACjournalVolNumPages{\mnras}{427}{}{127-145}.
\newblock
\begin{APACrefDOI} \doi{10.1111/j.1365-2966.2012.21948.x} \end{APACrefDOI}
\PrintBackRefs{\CurrentBib}

\bibitem [\protect \citeauthoryear {%
{Cappellari}%
}{%
{Cappellari}%
}{%
{\protect \APACyear {2017}}%
}]{%
cappellari17}
\APACinsertmetastar {%
cappellari17}%
\begin{APACrefauthors}%
{Cappellari}, M.%
\end{APACrefauthors}%
\unskip\
\newblock
\APACrefYearMonthDay{2017}{{\APACmonth{04}}}{},
\newblock
\unskip
\newblock
\APACjournalVolNumPages{\mnras}{466}{1}{798-811}.
\newblock
\begin{APACrefDOI} \doi{10.1093/mnras/stw3020} \end{APACrefDOI}
\PrintBackRefs{\CurrentBib}

\bibitem [\protect \citeauthoryear {%
{Cappellari}%
\ \BBA {} {Emsellem}%
}{%
{Cappellari}%
\ \BBA {} {Emsellem}%
}{%
{\protect \APACyear {2004}}%
}]{%
cappellari04}
\APACinsertmetastar {%
cappellari04}%
\begin{APACrefauthors}%
{Cappellari}, M.%
\BCBT {}\ \BBA {} {Emsellem}, E.%
\end{APACrefauthors}%
\unskip\
\newblock
\APACrefYearMonthDay{2004}{{\APACmonth{02}}}{},
\newblock
\unskip
\newblock
\APACjournalVolNumPages{\pasp}{116}{816}{138-147}.
\newblock
\begin{APACrefDOI} \doi{10.1086/381875} \end{APACrefDOI}
\PrintBackRefs{\CurrentBib}

\bibitem [\protect \citeauthoryear {%
{Cappellari}%
\ \protect \BOthers {.}}{%
{Cappellari}%
\ \protect \BOthers {.}}{%
{\protect \APACyear {2011}}%
}]{%
cappellari11}
\APACinsertmetastar {%
cappellari11}%
\begin{APACrefauthors}%
{Cappellari}, M.%
, {Emsellem}, E.%
, {Krajnovi{\'c}}, D.%
\ et al.\end{APACrefauthors}%
\unskip\
\newblock
\APACrefYearMonthDay{2011}{{\APACmonth{09}}}{},
\newblock
\unskip
\newblock
\APACjournalVolNumPages{\mnras}{416}{}{1680-1696}.
\newblock
\begin{APACrefDOI} \doi{10.1111/j.1365-2966.2011.18600.x} \end{APACrefDOI}
\PrintBackRefs{\CurrentBib}

\bibitem [\protect \citeauthoryear {%
{Chae}%
, {Bernardi}%
, {Dom{\'\i}nguez S{\'a}nchez}%
\BCBL {}\ \BBA {} {Sheth}%
}{%
{Chae}%
\ \protect \BOthers {.}}{%
{\protect \APACyear {2020}}%
}]{%
chae20}
\APACinsertmetastar {%
chae20}%
\begin{APACrefauthors}%
{Chae}, K\BHBI H.%
, {Bernardi}, M.%
, {Dom{\'\i}nguez S{\'a}nchez}, H.%
\BCBL {}\ \BBA {} {Sheth}, R\BPBI K.%
\end{APACrefauthors}%
\unskip\
\newblock
\APACrefYearMonthDay{2020}{{\APACmonth{11}}}{},
\newblock
\unskip
\newblock
\APACjournalVolNumPages{\apjl}{903}{2}{L31}.
\newblock
\begin{APACrefDOI} \doi{10.3847/2041-8213/abc2d3} \end{APACrefDOI}
\PrintBackRefs{\CurrentBib}

\bibitem [\protect \citeauthoryear {%
{Chan}%
, {Desai}%
\BCBL {}\ \BBA {} {Del Popolo}%
}{%
{Chan}%
\ \protect \BOthers {.}}{%
{\protect \APACyear {2022}}%
}]{%
chan22}
\APACinsertmetastar {%
chan22}%
\begin{APACrefauthors}%
{Chan}, M\BPBI H.%
, {Desai}, S.%
\BCBL {}\ \BBA {} {Del Popolo}, A.%
\end{APACrefauthors}%
\unskip\
\newblock
\APACrefYearMonthDay{2022}{{\APACmonth{05}}}{},
\newblock
\unskip
\newblock
\APACjournalVolNumPages{arXiv e-prints}{}{}{arXiv:2205.07515}.
\PrintBackRefs{\CurrentBib}

\bibitem [\protect \citeauthoryear {%
{Cherenkov Telescope Array Consortium}%
\ \protect \BOthers {.}}{%
{Cherenkov Telescope Array Consortium}%
\ \protect \BOthers {.}}{%
{\protect \APACyear {2019}}%
}]{%
acharya19}
\APACinsertmetastar {%
acharya19}%
\begin{APACrefauthors}%
{Cherenkov Telescope Array Consortium}%
, {Acharya}, B\BPBI S.%
, {Agudo}, I.%
\ et al.\end{APACrefauthors}%
\unskip\
\newblock
\APACrefYear{2019},
\newblock
\APACrefbtitle {{Science with the Cherenkov Telescope Array}} {{Science with
  the Cherenkov Telescope Array}}.
\newblock
\begin{APACrefDOI} \doi{10.1142/10986} \end{APACrefDOI}
\PrintBackRefs{\CurrentBib}

\bibitem [\protect \citeauthoryear {%
{de Lorenzi}%
\ \protect \BOthers {.}}{%
{de Lorenzi}%
\ \protect \BOthers {.}}{%
{\protect \APACyear {2009}}%
}]{%
delorenzi09}
\APACinsertmetastar {%
delorenzi09}%
\begin{APACrefauthors}%
{de Lorenzi}, F.%
, {Gerhard}, O.%
, {Coccato}, L.%
\ et al.\end{APACrefauthors}%
\unskip\
\newblock
\APACrefYearMonthDay{2009}{{\APACmonth{05}}}{},
\newblock
\unskip
\newblock
\APACjournalVolNumPages{\mnras}{395}{}{76-96}.
\newblock
\begin{APACrefDOI} \doi{10.1111/j.1365-2966.2009.14553.x} \end{APACrefDOI}
\PrintBackRefs{\CurrentBib}

\bibitem [\protect \citeauthoryear {%
{Deason}%
, {Belokurov}%
, {Evans}%
\BCBL {}\ \BBA {} {McCarthy}%
}{%
{Deason}%
\ \protect \BOthers {.}}{%
{\protect \APACyear {2012}}%
}]{%
deason12}
\APACinsertmetastar {%
deason12}%
\begin{APACrefauthors}%
{Deason}, A\BPBI J.%
, {Belokurov}, V.%
, {Evans}, N\BPBI W.%
\BCBL {}\ \BBA {} {McCarthy}, I\BPBI G.%
\end{APACrefauthors}%
\unskip\
\newblock
\APACrefYearMonthDay{2012}{{\APACmonth{03}}}{},
\newblock
\unskip
\newblock
\APACjournalVolNumPages{\apj}{748}{}{2}.
\newblock
\begin{APACrefDOI} \doi{10.1088/0004-637X/748/1/2} \end{APACrefDOI}
\PrintBackRefs{\CurrentBib}

\bibitem [\protect \citeauthoryear {%
{Di Valentino}%
, {Melchiorri}%
\BCBL {}\ \BBA {} {Silk}%
}{%
{Di Valentino}%
\ \protect \BOthers {.}}{%
{\protect \APACyear {2020}}%
}]{%
divalentino20}
\APACinsertmetastar {%
divalentino20}%
\begin{APACrefauthors}%
{Di Valentino}, E.%
, {Melchiorri}, A.%
\BCBL {}\ \BBA {} {Silk}, J.%
\end{APACrefauthors}%
\unskip\
\newblock
\APACrefYearMonthDay{2020}{{\APACmonth{02}}}{},
\newblock
\unskip
\newblock
\APACjournalVolNumPages{Nature Astronomy}{4}{}{196-203}.
\newblock
\begin{APACrefDOI} \doi{10.1038/s41550-019-0906-9} \end{APACrefDOI}
\PrintBackRefs{\CurrentBib}

\bibitem [\protect \citeauthoryear {%
{Dirsch}%
\ \protect \BOthers {.}}{%
{Dirsch}%
\ \protect \BOthers {.}}{%
{\protect \APACyear {2003}}%
}]{%
dirsch03}
\APACinsertmetastar {%
dirsch03}%
\begin{APACrefauthors}%
{Dirsch}, B.%
, {Richtler}, T.%
, {Geisler}, D.%
, {Forte}, J\BPBI C.%
, {Bassino}, L\BPBI P.%
\BCBL {}\ \BBA {} {Gieren}, W\BPBI P.%
\end{APACrefauthors}%
\unskip\
\newblock
\APACrefYearMonthDay{2003}{{\APACmonth{04}}}{},
\newblock
\unskip
\newblock
\APACjournalVolNumPages{\aj}{125}{}{1908-1925}.
\newblock
\begin{APACrefDOI} \doi{10.1086/368238} \end{APACrefDOI}
\PrintBackRefs{\CurrentBib}

\bibitem [\protect \citeauthoryear {%
{Evans}%
, {Wilkinson}%
, {Perrett}%
\BCBL {}\ \BBA {} {Bridges}%
}{%
{Evans}%
\ \protect \BOthers {.}}{%
{\protect \APACyear {2003}}%
}]{%
evans03}
\APACinsertmetastar {%
evans03}%
\begin{APACrefauthors}%
{Evans}, N\BPBI W.%
, {Wilkinson}, M\BPBI I.%
, {Perrett}, K\BPBI M.%
\BCBL {}\ \BBA {} {Bridges}, T\BPBI J.%
\end{APACrefauthors}%
\unskip\
\newblock
\APACrefYearMonthDay{2003}{{\APACmonth{02}}}{},
\newblock
\unskip
\newblock
\APACjournalVolNumPages{\apj}{583}{}{752-757}.
\newblock
\begin{APACrefDOI} \doi{10.1086/345400} \end{APACrefDOI}
\PrintBackRefs{\CurrentBib}

\bibitem [\protect \citeauthoryear {%
{Fahrion}%
\ \protect \BOthers {.}}{%
{Fahrion}%
\ \protect \BOthers {.}}{%
{\protect \APACyear {2019}}%
}]{%
fahrion19}
\APACinsertmetastar {%
fahrion19}%
\begin{APACrefauthors}%
{Fahrion}, K.%
, {Georgiev}, I.%
, {Hilker}, M.%
\ et al.\end{APACrefauthors}%
\unskip\
\newblock
\APACrefYearMonthDay{2019}{{\APACmonth{05}}}{},
\newblock
\unskip
\newblock
\APACjournalVolNumPages{\aap}{625}{}{A50}.
\newblock
\begin{APACrefDOI} \doi{10.1051/0004-6361/201834941} \end{APACrefDOI}
\PrintBackRefs{\CurrentBib}

\bibitem [\protect \citeauthoryear {%
{Famaey}%
, {Gentile}%
, {Bruneton}%
\BCBL {}\ \BBA {} {Zhao}%
}{%
{Famaey}%
\ \protect \BOthers {.}}{%
{\protect \APACyear {2007}}%
}]{%
famaey07}
\APACinsertmetastar {%
famaey07}%
\begin{APACrefauthors}%
{Famaey}, B.%
, {Gentile}, G.%
, {Bruneton}, J\BHBI P.%
\BCBL {}\ \BBA {} {Zhao}, H.%
\end{APACrefauthors}%
\unskip\
\newblock
\APACrefYearMonthDay{2007}{{\APACmonth{03}}}{},
\newblock
\unskip
\newblock
\APACjournalVolNumPages{\prd}{75}{6}{063002}.
\newblock
\begin{APACrefDOI} \doi{10.1103/PhysRevD.75.063002} \end{APACrefDOI}
\PrintBackRefs{\CurrentBib}

\bibitem [\protect \citeauthoryear {%
{Famaey}%
\ \BBA {} {McGaugh}%
}{%
{Famaey}%
\ \BBA {} {McGaugh}%
}{%
{\protect \APACyear {2013}}%
}]{%
famaey13}
\APACinsertmetastar {%
famaey13}%
\begin{APACrefauthors}%
{Famaey}, B.%
\BCBT {}\ \BBA {} {McGaugh}, S.%
\end{APACrefauthors}%
\unskip\
\newblock
\APACrefYearMonthDay{2013}{{\APACmonth{10}}}{},
\newblock
\unskip
\newblock
\APACjournalVolNumPages{ArXiv e-prints}{}{}{}.
\PrintBackRefs{\CurrentBib}

\bibitem [\protect \citeauthoryear {%
{Famaey}%
\ \BBA {} {McGaugh}%
}{%
{Famaey}%
\ \BBA {} {McGaugh}%
}{%
{\protect \APACyear {2012}}%
}]{%
famaey12}
\APACinsertmetastar {%
famaey12}%
\begin{APACrefauthors}%
{Famaey}, B.%
\BCBT {}\ \BBA {} {McGaugh}, S\BPBI S.%
\end{APACrefauthors}%
\unskip\
\newblock
\APACrefYearMonthDay{2012}{{\APACmonth{09}}}{},
\newblock
\unskip
\newblock
\APACjournalVolNumPages{Living Reviews in Relativity}{15}{1}{10}.
\newblock
\begin{APACrefDOI} \doi{10.12942/lrr-2012-10} \end{APACrefDOI}
\PrintBackRefs{\CurrentBib}

\bibitem [\protect \citeauthoryear {%
{Gerhard}%
, {Kronawitter}%
, {Saglia}%
\BCBL {}\ \BBA {} {Bender}%
}{%
{Gerhard}%
\ \protect \BOthers {.}}{%
{\protect \APACyear {2001}}%
}]{%
gerhard01}
\APACinsertmetastar {%
gerhard01}%
\begin{APACrefauthors}%
{Gerhard}, O.%
, {Kronawitter}, A.%
, {Saglia}, R\BPBI P.%
\BCBL {}\ \BBA {} {Bender}, R.%
\end{APACrefauthors}%
\unskip\
\newblock
\APACrefYearMonthDay{2001}{{\APACmonth{04}}}{},
\newblock
\unskip
\newblock
\APACjournalVolNumPages{\aj}{121}{}{1936-1951}.
\newblock
\begin{APACrefDOI} \doi{10.1086/319940} \end{APACrefDOI}
\PrintBackRefs{\CurrentBib}

\bibitem [\protect \citeauthoryear {%
{Gibbons}%
\ \BBA {} {Hawking}%
}{%
{Gibbons}%
\ \BBA {} {Hawking}%
}{%
{\protect \APACyear {1977}}%
}]{%
gibbons77}
\APACinsertmetastar {%
gibbons77}%
\begin{APACrefauthors}%
{Gibbons}, G\BPBI W.%
\BCBT {}\ \BBA {} {Hawking}, S\BPBI W.%
\end{APACrefauthors}%
\unskip\
\newblock
\APACrefYearMonthDay{1977}{{\APACmonth{05}}}{},
\newblock
\unskip
\newblock
\APACjournalVolNumPages{\prd}{15}{}{2738-2751}.
\newblock
\begin{APACrefDOI} \doi{10.1103/PhysRevD.15.2738} \end{APACrefDOI}
\PrintBackRefs{\CurrentBib}

\bibitem [\protect \citeauthoryear {%
{Hansen}%
\ \BBA {} {Moore}%
}{%
{Hansen}%
\ \BBA {} {Moore}%
}{%
{\protect \APACyear {2006}}%
}]{%
hansen06}
\APACinsertmetastar {%
hansen06}%
\begin{APACrefauthors}%
{Hansen}, S\BPBI H.%
\BCBT {}\ \BBA {} {Moore}, B.%
\end{APACrefauthors}%
\unskip\
\newblock
\APACrefYearMonthDay{2006}{{\APACmonth{03}}}{},
\newblock
\unskip
\newblock
\APACjournalVolNumPages{\na}{11}{}{333-338}.
\newblock
\begin{APACrefDOI} \doi{10.1016/j.newast.2005.09.001} \end{APACrefDOI}
\PrintBackRefs{\CurrentBib}

\bibitem [\protect \citeauthoryear {%
{Harris}%
\ \BBA {} {Canterna}%
}{%
{Harris}%
\ \BBA {} {Canterna}%
}{%
{\protect \APACyear {1977}}%
}]{%
harris77}
\APACinsertmetastar {%
harris77}%
\begin{APACrefauthors}%
{Harris}, H\BPBI C.%
\BCBT {}\ \BBA {} {Canterna}, R.%
\end{APACrefauthors}%
\unskip\
\newblock
\APACrefYearMonthDay{1977}{{\APACmonth{10}}}{},
\newblock
\unskip
\newblock
\APACjournalVolNumPages{\aj}{82}{}{798-804}.
\newblock
\begin{APACrefDOI} \doi{10.1086/112129} \end{APACrefDOI}
\PrintBackRefs{\CurrentBib}

\bibitem [\protect \citeauthoryear {%
{Hawking}%
}{%
{Hawking}%
}{%
{\protect \APACyear {1975}}%
}]{%
hawking75}
\APACinsertmetastar {%
hawking75}%
\begin{APACrefauthors}%
{Hawking}, S\BPBI W.%
\end{APACrefauthors}%
\unskip\
\newblock
\APACrefYearMonthDay{1975}{{\APACmonth{08}}}{},
\newblock
\unskip
\newblock
\APACjournalVolNumPages{Communications in Mathematical Physics}{43}{}{199-220}.
\newblock
\begin{APACrefDOI} \doi{10.1007/BF02345020} \end{APACrefDOI}
\PrintBackRefs{\CurrentBib}

\bibitem [\protect \citeauthoryear {%
{Heinesen}%
\ \BBA {} {Buchert}%
}{%
{Heinesen}%
\ \BBA {} {Buchert}%
}{%
{\protect \APACyear {2020}}%
}]{%
heinesen20}
\APACinsertmetastar {%
heinesen20}%
\begin{APACrefauthors}%
{Heinesen}, A.%
\BCBT {}\ \BBA {} {Buchert}, T.%
\end{APACrefauthors}%
\unskip\
\newblock
\APACrefYearMonthDay{2020}{{\APACmonth{08}}}{},
\newblock
\unskip
\newblock
\APACjournalVolNumPages{Classical and Quantum Gravity}{37}{16}{164001}.
\newblock
\begin{APACrefDOI} \doi{10.1088/1361-6382/ab954b} \end{APACrefDOI}
\PrintBackRefs{\CurrentBib}

\bibitem [\protect \citeauthoryear {%
{Heisler}%
, {Tremaine}%
\BCBL {}\ \BBA {} {Bahcall}%
}{%
{Heisler}%
\ \protect \BOthers {.}}{%
{\protect \APACyear {1985}}%
}]{%
heisler85}
\APACinsertmetastar {%
heisler85}%
\begin{APACrefauthors}%
{Heisler}, J.%
, {Tremaine}, S.%
\BCBL {}\ \BBA {} {Bahcall}, J\BPBI N.%
\end{APACrefauthors}%
\unskip\
\newblock
\APACrefYearMonthDay{1985}{{\APACmonth{11}}}{},
\newblock
\unskip
\newblock
\APACjournalVolNumPages{\apj}{298}{}{8-17}.
\newblock
\begin{APACrefDOI} \doi{10.1086/163584} \end{APACrefDOI}
\PrintBackRefs{\CurrentBib}

\bibitem [\protect \citeauthoryear {%
{Hilker}%
\ \protect \BOthers {.}}{%
{Hilker}%
\ \protect \BOthers {.}}{%
{\protect \APACyear {2018}}%
}]{%
hilker18}
\APACinsertmetastar {%
hilker18}%
\begin{APACrefauthors}%
{Hilker}, M.%
, {Richtler}, T.%
, {Barbosa}, C\BPBI E.%
, {Arnaboldi}, M.%
, {Coccato}, L.%
\BCBL {}\ \BBA {} {Mendes de Oliveira}, C.%
\end{APACrefauthors}%
\unskip\
\newblock
\APACrefYearMonthDay{2018}{{\APACmonth{11}}}{},
\newblock
\unskip
\newblock
\APACjournalVolNumPages{\aap}{619}{}{A70}.
\newblock
\begin{APACrefDOI} \doi{10.1051/0004-6361/201731737} \end{APACrefDOI}
\PrintBackRefs{\CurrentBib}

\bibitem [\protect \citeauthoryear {%
{Hossenfelder}%
}{%
{Hossenfelder}%
}{%
{\protect \APACyear {2017}}%
}]{%
hossenfelder17}
\APACinsertmetastar {%
hossenfelder17}%
\begin{APACrefauthors}%
{Hossenfelder}, S.%
\end{APACrefauthors}%
\unskip\
\newblock
\APACrefYearMonthDay{2017}{{\APACmonth{06}}}{},
\newblock
\unskip
\newblock
\APACjournalVolNumPages{\prd}{95}{12}{124018}.
\newblock
\begin{APACrefDOI} \doi{10.1103/PhysRevD.95.124018} \end{APACrefDOI}
\PrintBackRefs{\CurrentBib}

\bibitem [\protect \citeauthoryear {%
{Hossenfelder}%
\ \BBA {} {Mistele}%
}{%
{Hossenfelder}%
\ \BBA {} {Mistele}%
}{%
{\protect \APACyear {2018}}%
}]{%
hossenfelder18}
\APACinsertmetastar {%
hossenfelder18}%
\begin{APACrefauthors}%
{Hossenfelder}, S.%
\BCBT {}\ \BBA {} {Mistele}, T.%
\end{APACrefauthors}%
\unskip\
\newblock
\APACrefYearMonthDay{2018}{{\APACmonth{01}}}{},
\newblock
\unskip
\newblock
\APACjournalVolNumPages{International Journal of Modern Physics
  D}{27}{14}{1847010}.
\newblock
\begin{APACrefDOI} \doi{10.1142/S0218271818470107} \end{APACrefDOI}
\PrintBackRefs{\CurrentBib}

\bibitem [\protect \citeauthoryear {%
{Humphrey}%
, {Buote}%
, {Canizares}%
, {Fabian}%
\BCBL {}\ \BBA {} {Miller}%
}{%
{Humphrey}%
\ \protect \BOthers {.}}{%
{\protect \APACyear {2011}}%
}]{%
humphrey11}
\APACinsertmetastar {%
humphrey11}%
\begin{APACrefauthors}%
{Humphrey}, P\BPBI J.%
, {Buote}, D\BPBI A.%
, {Canizares}, C\BPBI R.%
, {Fabian}, A\BPBI C.%
\BCBL {}\ \BBA {} {Miller}, J\BPBI M.%
\end{APACrefauthors}%
\unskip\
\newblock
\APACrefYearMonthDay{2011}{{\APACmonth{03}}}{},
\newblock
\unskip
\newblock
\APACjournalVolNumPages{\apj}{729}{1}{53}.
\newblock
\begin{APACrefDOI} \doi{10.1088/0004-637X/729/1/53} \end{APACrefDOI}
\PrintBackRefs{\CurrentBib}

\bibitem [\protect \citeauthoryear {%
{Humphrey}%
, {Buote}%
, {O'Sullivan}%
\BCBL {}\ \BBA {} {Ponman}%
}{%
{Humphrey}%
\ \protect \BOthers {.}}{%
{\protect \APACyear {2012}}%
}]{%
humphrey12}
\APACinsertmetastar {%
humphrey12}%
\begin{APACrefauthors}%
{Humphrey}, P\BPBI J.%
, {Buote}, D\BPBI A.%
, {O'Sullivan}, E.%
\BCBL {}\ \BBA {} {Ponman}, T\BPBI J.%
\end{APACrefauthors}%
\unskip\
\newblock
\APACrefYearMonthDay{2012}{{\APACmonth{08}}}{},
\newblock
\unskip
\newblock
\APACjournalVolNumPages{\apj}{755}{2}{166}.
\newblock
\begin{APACrefDOI} \doi{10.1088/0004-637X/755/2/166} \end{APACrefDOI}
\PrintBackRefs{\CurrentBib}

\bibitem [\protect \citeauthoryear {%
{Jacobson}%
}{%
{Jacobson}%
}{%
{\protect \APACyear {1995}}%
}]{%
jacobson95}
\APACinsertmetastar {%
jacobson95}%
\begin{APACrefauthors}%
{Jacobson}, T.%
\end{APACrefauthors}%
\unskip\
\newblock
\APACrefYearMonthDay{1995}{{\APACmonth{08}}}{},
\newblock
\unskip
\newblock
\APACjournalVolNumPages{Physical Review Letters}{75}{}{1260-1263}.
\newblock
\begin{APACrefDOI} \doi{10.1103/PhysRevLett.75.1260} \end{APACrefDOI}
\PrintBackRefs{\CurrentBib}

\bibitem [\protect \citeauthoryear {%
{Kroupa}%
}{%
{Kroupa}%
}{%
{\protect \APACyear {2015}}%
}]{%
kroupa15}
\APACinsertmetastar {%
kroupa15}%
\begin{APACrefauthors}%
{Kroupa}, P.%
\end{APACrefauthors}%
\unskip\
\newblock
\APACrefYearMonthDay{2015}{{\APACmonth{02}}}{},
\newblock
\unskip
\newblock
\APACjournalVolNumPages{Canadian Journal of Physics}{93}{}{169-202}.
\newblock
\begin{APACrefDOI} \doi{10.1139/cjp-2014-0179} \end{APACrefDOI}
\PrintBackRefs{\CurrentBib}

\bibitem [\protect \citeauthoryear {%
{Kurapati}%
, {Chengalur}%
, {Pustilnik}%
\BCBL {}\ \BBA {} {Kamphuis}%
}{%
{Kurapati}%
\ \protect \BOthers {.}}{%
{\protect \APACyear {2018}}%
}]{%
kurapati18}
\APACinsertmetastar {%
kurapati18}%
\begin{APACrefauthors}%
{Kurapati}, S.%
, {Chengalur}, J\BPBI N.%
, {Pustilnik}, S.%
\BCBL {}\ \BBA {} {Kamphuis}, P.%
\end{APACrefauthors}%
\unskip\
\newblock
\APACrefYearMonthDay{2018}{{\APACmonth{09}}}{},
\newblock
\unskip
\newblock
\APACjournalVolNumPages{\mnras}{479}{}{228-239}.
\newblock
\begin{APACrefDOI} \doi{10.1093/mnras/sty1397} \end{APACrefDOI}
\PrintBackRefs{\CurrentBib}

\bibitem [\protect \citeauthoryear {%
{Lane}%
, {Salinas}%
\BCBL {}\ \BBA {} {Richtler}%
}{%
{Lane}%
\ \protect \BOthers {.}}{%
{\protect \APACyear {2013}}%
}]{%
lane13}
\APACinsertmetastar {%
lane13}%
\begin{APACrefauthors}%
{Lane}, R\BPBI R.%
, {Salinas}, R.%
\BCBL {}\ \BBA {} {Richtler}, T.%
\end{APACrefauthors}%
\unskip\
\newblock
\APACrefYearMonthDay{2013}{{\APACmonth{01}}}{},
\newblock
\unskip
\newblock
\APACjournalVolNumPages{\aap}{549}{}{A148}.
\newblock
\begin{APACrefDOI} \doi{10.1051/0004-6361/201220231} \end{APACrefDOI}
\PrintBackRefs{\CurrentBib}

\bibitem [\protect \citeauthoryear {%
{Lane}%
, {Salinas}%
\BCBL {}\ \BBA {} {Richtler}%
}{%
{Lane}%
\ \protect \BOthers {.}}{%
{\protect \APACyear {2015}}%
}]{%
lane15}
\APACinsertmetastar {%
lane15}%
\begin{APACrefauthors}%
{Lane}, R\BPBI R.%
, {Salinas}, R.%
\BCBL {}\ \BBA {} {Richtler}, T.%
\end{APACrefauthors}%
\unskip\
\newblock
\APACrefYearMonthDay{2015}{{\APACmonth{02}}}{},
\newblock
\unskip
\newblock
\APACjournalVolNumPages{\aap}{574}{}{A93}.
\newblock
\begin{APACrefDOI} \doi{10.1051/0004-6361/201424074} \end{APACrefDOI}
\PrintBackRefs{\CurrentBib}

\bibitem [\protect \citeauthoryear {%
{Lelli}%
, {McGaugh}%
, {Schombert}%
, {Desmond}%
\BCBL {}\ \BBA {} {Katz}%
}{%
{Lelli}%
\ \protect \BOthers {.}}{%
{\protect \APACyear {2019}}%
}]{%
lelli19}
\APACinsertmetastar {%
lelli19}%
\begin{APACrefauthors}%
{Lelli}, F.%
, {McGaugh}, S\BPBI S.%
, {Schombert}, J\BPBI M.%
, {Desmond}, H.%
\BCBL {}\ \BBA {} {Katz}, H.%
\end{APACrefauthors}%
\unskip\
\newblock
\APACrefYearMonthDay{2019}{{\APACmonth{04}}}{},
\newblock
\unskip
\newblock
\APACjournalVolNumPages{\mnras}{484}{3}{3267-3278}.
\newblock
\begin{APACrefDOI} \doi{10.1093/mnras/stz205} \end{APACrefDOI}
\PrintBackRefs{\CurrentBib}

\bibitem [\protect \citeauthoryear {%
{Lelli}%
, {McGaugh}%
, {Schombert}%
\BCBL {}\ \BBA {} {Pawlowski}%
}{%
{Lelli}%
\ \protect \BOthers {.}}{%
{\protect \APACyear {2017}}%
}]{%
lelli17}
\APACinsertmetastar {%
lelli17}%
\begin{APACrefauthors}%
{Lelli}, F.%
, {McGaugh}, S\BPBI S.%
, {Schombert}, J\BPBI M.%
\BCBL {}\ \BBA {} {Pawlowski}, M\BPBI S.%
\end{APACrefauthors}%
\unskip\
\newblock
\APACrefYearMonthDay{2017}{{\APACmonth{02}}}{},
\newblock
\unskip
\newblock
\APACjournalVolNumPages{\apj}{836}{2}{152}.
\newblock
\begin{APACrefDOI} \doi{10.3847/1538-4357/836/2/152} \end{APACrefDOI}
\PrintBackRefs{\CurrentBib}

\bibitem [\protect \citeauthoryear {%
{Li}%
, {Lelli}%
, {McGaugh}%
, {Schombert}%
\BCBL {}\ \BBA {} {Chae}%
}{%
{Li}%
\ \protect \BOthers {.}}{%
{\protect \APACyear {2021}}%
}]{%
li21}
\APACinsertmetastar {%
li21}%
\begin{APACrefauthors}%
{Li}, P.%
, {Lelli}, F.%
, {McGaugh}, S.%
, {Schombert}, J.%
\BCBL {}\ \BBA {} {Chae}, K\BHBI H.%
\end{APACrefauthors}%
\unskip\
\newblock
\APACrefYearMonthDay{2021}{{\APACmonth{02}}}{},
\newblock
\unskip
\newblock
\APACjournalVolNumPages{\aap}{646}{}{L13}.
\newblock
\begin{APACrefDOI} \doi{10.1051/0004-6361/202040101} \end{APACrefDOI}
\PrintBackRefs{\CurrentBib}

\bibitem [\protect \citeauthoryear {%
{Lovell}%
\ \protect \BOthers {.}}{%
{Lovell}%
\ \protect \BOthers {.}}{%
{\protect \APACyear {2018}}%
}]{%
lovell18}
\APACinsertmetastar {%
lovell18}%
\begin{APACrefauthors}%
{Lovell}, M\BPBI R.%
, {Pillepich}, A.%
, {Genel}, S.%
\ et al.\end{APACrefauthors}%
\unskip\
\newblock
\APACrefYearMonthDay{2018}{{\APACmonth{12}}}{},
\newblock
\unskip
\newblock
\APACjournalVolNumPages{\mnras}{481}{2}{1950-1975}.
\newblock
\begin{APACrefDOI} \doi{10.1093/mnras/sty2339} \end{APACrefDOI}
\PrintBackRefs{\CurrentBib}

\bibitem [\protect \citeauthoryear {%
{Magorrian}%
\ \BBA {} {Ballantyne}%
}{%
{Magorrian}%
\ \BBA {} {Ballantyne}%
}{%
{\protect \APACyear {2001}}%
}]{%
magorrian01}
\APACinsertmetastar {%
magorrian01}%
\begin{APACrefauthors}%
{Magorrian}, J.%
\BCBT {}\ \BBA {} {Ballantyne}, D.%
\end{APACrefauthors}%
\unskip\
\newblock
\APACrefYearMonthDay{2001}{{\APACmonth{04}}}{},
\newblock
\unskip
\newblock
\APACjournalVolNumPages{\mnras}{322}{}{702-714}.
\newblock
\begin{APACrefDOI} \doi{10.1046/j.1365-8711.2001.04150.x} \end{APACrefDOI}
\PrintBackRefs{\CurrentBib}

\bibitem [\protect \citeauthoryear {%
{Mamon}%
\ \BBA {} {{\L}okas}%
}{%
{Mamon}%
\ \BBA {} {{\L}okas}%
}{%
{\protect \APACyear {2005}}%
}]{%
mamon05}
\APACinsertmetastar {%
mamon05}%
\begin{APACrefauthors}%
{Mamon}, G\BPBI A.%
\BCBT {}\ \BBA {} {{\L}okas}, E\BPBI L.%
\end{APACrefauthors}%
\unskip\
\newblock
\APACrefYearMonthDay{2005}{{\APACmonth{11}}}{},
\newblock
\unskip
\newblock
\APACjournalVolNumPages{\mnras}{363}{}{705-722}.
\newblock
\begin{APACrefDOI} \doi{10.1111/j.1365-2966.2005.09400.x} \end{APACrefDOI}
\PrintBackRefs{\CurrentBib}

\bibitem [\protect \citeauthoryear {%
{McGaugh}%
}{%
{McGaugh}%
}{%
{\protect \APACyear {2005}}%
}]{%
mcgaugh05}
\APACinsertmetastar {%
mcgaugh05}%
\begin{APACrefauthors}%
{McGaugh}, S\BPBI S.%
\end{APACrefauthors}%
\unskip\
\newblock
\APACrefYearMonthDay{2005}{{\APACmonth{10}}}{},
\newblock
\unskip
\newblock
\APACjournalVolNumPages{\apj}{632}{}{859-871}.
\newblock
\begin{APACrefDOI} \doi{10.1086/432968} \end{APACrefDOI}
\PrintBackRefs{\CurrentBib}

\bibitem [\protect \citeauthoryear {%
{McGaugh}%
}{%
{McGaugh}%
}{%
{\protect \APACyear {2012}}%
}]{%
mcgaugh12}
\APACinsertmetastar {%
mcgaugh12}%
\begin{APACrefauthors}%
{McGaugh}, S\BPBI S.%
\end{APACrefauthors}%
\unskip\
\newblock
\APACrefYearMonthDay{2012}{{\APACmonth{02}}}{},
\newblock
\unskip
\newblock
\APACjournalVolNumPages{\aj}{143}{}{40}.
\newblock
\begin{APACrefDOI} \doi{10.1088/0004-6256/143/2/40} \end{APACrefDOI}
\PrintBackRefs{\CurrentBib}

\bibitem [\protect \citeauthoryear {%
{McGaugh}%
, {Lelli}%
\BCBL {}\ \BBA {} {Schombert}%
}{%
{McGaugh}%
\ \protect \BOthers {.}}{%
{\protect \APACyear {2016}}%
}]{%
mcgaugh16}
\APACinsertmetastar {%
mcgaugh16}%
\begin{APACrefauthors}%
{McGaugh}, S\BPBI S.%
, {Lelli}, F.%
\BCBL {}\ \BBA {} {Schombert}, J\BPBI M.%
\end{APACrefauthors}%
\unskip\
\newblock
\APACrefYearMonthDay{2016}{{\APACmonth{11}}}{},
\newblock
\unskip
\newblock
\APACjournalVolNumPages{Physical Review Letters}{117}{20}{201101}.
\newblock
\begin{APACrefDOI} \doi{10.1103/PhysRevLett.117.201101} \end{APACrefDOI}
\PrintBackRefs{\CurrentBib}

\bibitem [\protect \citeauthoryear {%
{M{\'e}ndez}%
, {Teodorescu}%
, {Kudritzki}%
\BCBL {}\ \BBA {} {Burkert}%
}{%
{M{\'e}ndez}%
\ \protect \BOthers {.}}{%
{\protect \APACyear {2009}}%
}]{%
mendez09}
\APACinsertmetastar {%
mendez09}%
\begin{APACrefauthors}%
{M{\'e}ndez}, R\BPBI H.%
, {Teodorescu}, A\BPBI M.%
, {Kudritzki}, R\BHBI P.%
\BCBL {}\ \BBA {} {Burkert}, A.%
\end{APACrefauthors}%
\unskip\
\newblock
\APACrefYearMonthDay{2009}{{\APACmonth{01}}}{},
\newblock
\unskip
\newblock
\APACjournalVolNumPages{\apj}{691}{}{228-240}.
\newblock
\begin{APACrefDOI} \doi{10.1088/0004-637X/691/1/228} \end{APACrefDOI}
\PrintBackRefs{\CurrentBib}

\bibitem [\protect \citeauthoryear {%
{Milgrom}%
}{%
{Milgrom}%
}{%
{\protect \APACyear {1983}}%
{\protect \APACexlab {{\protect \BCnt {1}}}}}]{%
milgrom83a}
\APACinsertmetastar {%
milgrom83a}%
\begin{APACrefauthors}%
{Milgrom}, M.%
\end{APACrefauthors}%
\unskip\
\newblock
\APACrefYearMonthDay{1983{\protect \BCnt {1}}}{{\APACmonth{07}}}{},
\newblock
\unskip
\newblock
\APACjournalVolNumPages{\apj}{270}{}{365-370}.
\newblock
\begin{APACrefDOI} \doi{10.1086/161130} \end{APACrefDOI}
\PrintBackRefs{\CurrentBib}

\bibitem [\protect \citeauthoryear {%
{Milgrom}%
}{%
{Milgrom}%
}{%
{\protect \APACyear {1983}}%
{\protect \APACexlab {{\protect \BCnt {2}}}}}]{%
milgrom83b}
\APACinsertmetastar {%
milgrom83b}%
\begin{APACrefauthors}%
{Milgrom}, M.%
\end{APACrefauthors}%
\unskip\
\newblock
\APACrefYearMonthDay{1983{\protect \BCnt {2}}}{{\APACmonth{07}}}{},
\newblock
\unskip
\newblock
\APACjournalVolNumPages{\apj}{270}{}{371-383}.
\newblock
\begin{APACrefDOI} \doi{10.1086/161131} \end{APACrefDOI}
\PrintBackRefs{\CurrentBib}

\bibitem [\protect \citeauthoryear {%
{Milgrom}%
}{%
{Milgrom}%
}{%
{\protect \APACyear {1983}}%
{\protect \APACexlab {{\protect \BCnt {3}}}}}]{%
milgrom83c}
\APACinsertmetastar {%
milgrom83c}%
\begin{APACrefauthors}%
{Milgrom}, M.%
\end{APACrefauthors}%
\unskip\
\newblock
\APACrefYearMonthDay{1983{\protect \BCnt {3}}}{{\APACmonth{07}}}{},
\newblock
\unskip
\newblock
\APACjournalVolNumPages{\apj}{270}{}{384-389}.
\newblock
\begin{APACrefDOI} \doi{10.1086/161132} \end{APACrefDOI}
\PrintBackRefs{\CurrentBib}

\bibitem [\protect \citeauthoryear {%
{Milgrom}%
}{%
{Milgrom}%
}{%
{\protect \APACyear {2012}}%
}]{%
milgrom12}
\APACinsertmetastar {%
milgrom12}%
\begin{APACrefauthors}%
{Milgrom}, M.%
\end{APACrefauthors}%
\unskip\
\newblock
\APACrefYearMonthDay{2012}{{\APACmonth{09}}}{},
\newblock
\unskip
\newblock
\APACjournalVolNumPages{Physical Review Letters}{109}{13}{131101}.
\newblock
\begin{APACrefDOI} \doi{10.1103/PhysRevLett.109.131101} \end{APACrefDOI}
\PrintBackRefs{\CurrentBib}

\bibitem [\protect \citeauthoryear {%
{Milgrom}%
}{%
{Milgrom}%
}{%
{\protect \APACyear {2023}}%
}]{%
milgrom23}
\APACinsertmetastar {%
milgrom23}%
\begin{APACrefauthors}%
{Milgrom}, M.%
\end{APACrefauthors}%
\unskip\
\newblock
\APACrefYearMonthDay{2023}{{\APACmonth{09}}}{},
\newblock
\unskip
\newblock
\APACjournalVolNumPages{\prd}{108}{6}{063009}.
\newblock
\begin{APACrefDOI} \doi{10.1103/PhysRevD.108.063009} \end{APACrefDOI}
\PrintBackRefs{\CurrentBib}

\bibitem [\protect \citeauthoryear {%
{Morganti}%
, {Gerhard}%
, {Coccato}%
, {Martinez-Valpuesta}%
\BCBL {}\ \BBA {} {Arnaboldi}%
}{%
{Morganti}%
\ \protect \BOthers {.}}{%
{\protect \APACyear {2013}}%
}]{%
morganti13}
\APACinsertmetastar {%
morganti13}%
\begin{APACrefauthors}%
{Morganti}, L.%
, {Gerhard}, O.%
, {Coccato}, L.%
, {Martinez-Valpuesta}, I.%
\BCBL {}\ \BBA {} {Arnaboldi}, M.%
\end{APACrefauthors}%
\unskip\
\newblock
\APACrefYearMonthDay{2013}{{\APACmonth{06}}}{},
\newblock
\unskip
\newblock
\APACjournalVolNumPages{\mnras}{431}{}{3570-3588}.
\newblock
\begin{APACrefDOI} \doi{10.1093/mnras/stt442} \end{APACrefDOI}
\PrintBackRefs{\CurrentBib}

\bibitem [\protect \citeauthoryear {%
{Niemi}%
, {Hein{\"a}m{\"a}ki}%
, {Nurmi}%
\BCBL {}\ \BBA {} {Saar}%
}{%
{Niemi}%
\ \protect \BOthers {.}}{%
{\protect \APACyear {2010}}%
}]{%
niemi10}
\APACinsertmetastar {%
niemi10}%
\begin{APACrefauthors}%
{Niemi}, S\BHBI M.%
, {Hein{\"a}m{\"a}ki}, P.%
, {Nurmi}, P.%
\BCBL {}\ \BBA {} {Saar}, E.%
\end{APACrefauthors}%
\unskip\
\newblock
\APACrefYearMonthDay{2010}{{\APACmonth{06}}}{},
\newblock
\unskip
\newblock
\APACjournalVolNumPages{\mnras}{405}{}{477-493}.
\newblock
\begin{APACrefDOI} \doi{10.1111/j.1365-2966.2010.16457.x} \end{APACrefDOI}
\PrintBackRefs{\CurrentBib}

\bibitem [\protect \citeauthoryear {%
{Ostriker}%
\ \BBA {} {Steinhardt}%
}{%
{Ostriker}%
\ \BBA {} {Steinhardt}%
}{%
{\protect \APACyear {1995}}%
}]{%
ostriker95}
\APACinsertmetastar {%
ostriker95}%
\begin{APACrefauthors}%
{Ostriker}, J\BPBI P.%
\BCBT {}\ \BBA {} {Steinhardt}, P\BPBI J.%
\end{APACrefauthors}%
\unskip\
\newblock
\APACrefYearMonthDay{1995}{{\APACmonth{10}}}{},
\newblock
\unskip
\newblock
\APACjournalVolNumPages{\nat}{377}{}{600-602}.
\newblock
\begin{APACrefDOI} \doi{10.1038/377600a0} \end{APACrefDOI}
\PrintBackRefs{\CurrentBib}

\bibitem [\protect \citeauthoryear {%
{O'Sullivan}%
, {Sanderson}%
\BCBL {}\ \BBA {} {Ponman}%
}{%
{O'Sullivan}%
\ \protect \BOthers {.}}{%
{\protect \APACyear {2007}}%
}]{%
osullivan07}
\APACinsertmetastar {%
osullivan07}%
\begin{APACrefauthors}%
{O'Sullivan}, E.%
, {Sanderson}, A\BPBI J\BPBI R.%
\BCBL {}\ \BBA {} {Ponman}, T\BPBI J.%
\end{APACrefauthors}%
\unskip\
\newblock
\APACrefYearMonthDay{2007}{{\APACmonth{10}}}{},
\newblock
\unskip
\newblock
\APACjournalVolNumPages{\mnras}{380}{4}{1409-1421}.
\newblock
\begin{APACrefDOI} \doi{10.1111/j.1365-2966.2007.12229.x} \end{APACrefDOI}
\PrintBackRefs{\CurrentBib}

\bibitem [\protect \citeauthoryear {%
{Padmanabhan}%
}{%
{Padmanabhan}%
}{%
{\protect \APACyear {2015}}%
}]{%
padmanabhan15}
\APACinsertmetastar {%
padmanabhan15}%
\begin{APACrefauthors}%
{Padmanabhan}, T.%
\end{APACrefauthors}%
\unskip\
\newblock
\APACrefYearMonthDay{2015}{{\APACmonth{02}}}{},
\newblock
\unskip
\newblock
\APACjournalVolNumPages{Modern Physics Letters A}{30}{}{1540007}.
\newblock
\begin{APACrefDOI} \doi{10.1142/S0217732315400076} \end{APACrefDOI}
\PrintBackRefs{\CurrentBib}

\bibitem [\protect \citeauthoryear {%
{Padmanabhan}%
}{%
{Padmanabhan}%
}{%
{\protect \APACyear {2018}}%
}]{%
padmanabhan18}
\APACinsertmetastar {%
padmanabhan18}%
\begin{APACrefauthors}%
{Padmanabhan}, T.%
\end{APACrefauthors}%
\unskip\
\newblock
\APACrefYearMonthDay{2018}{{\APACmonth{01}}}{},
\newblock
\unskip
\newblock
\APACjournalVolNumPages{International Journal of Modern Physics
  D}{27}{14}{1846004}.
\newblock
\begin{APACrefDOI} \doi{10.1142/S0218271818460045} \end{APACrefDOI}
\PrintBackRefs{\CurrentBib}

\bibitem [\protect \citeauthoryear {%
{Planck Collaboration}%
\ \protect \BOthers {.}}{%
{Planck Collaboration}%
\ \protect \BOthers {.}}{%
{\protect \APACyear {2014}}%
}]{%
planckcoll14}
\APACinsertmetastar {%
planckcoll14}%
\begin{APACrefauthors}%
{Planck Collaboration}%
, {Ade}, P\BPBI A\BPBI R.%
, {Aghanim}, N.%
\ et al.\end{APACrefauthors}%
\unskip\
\newblock
\APACrefYearMonthDay{2014}{{\APACmonth{11}}}{},
\newblock
\unskip
\newblock
\APACjournalVolNumPages{\aap}{571}{}{A16}.
\newblock
\begin{APACrefDOI} \doi{10.1051/0004-6361/201321591} \end{APACrefDOI}
\PrintBackRefs{\CurrentBib}

\bibitem [\protect \citeauthoryear {%
{Pointecouteau}%
\ \BBA {} {Silk}%
}{%
{Pointecouteau}%
\ \BBA {} {Silk}%
}{%
{\protect \APACyear {2005}}%
}]{%
pointecouteau05}
\APACinsertmetastar {%
pointecouteau05}%
\begin{APACrefauthors}%
{Pointecouteau}, E.%
\BCBT {}\ \BBA {} {Silk}, J.%
\end{APACrefauthors}%
\unskip\
\newblock
\APACrefYearMonthDay{2005}{{\APACmonth{12}}}{},
\newblock
\unskip
\newblock
\APACjournalVolNumPages{\mnras}{364}{}{654-658}.
\newblock
\begin{APACrefDOI} \doi{10.1111/j.1365-2966.2005.09590.x} \end{APACrefDOI}
\PrintBackRefs{\CurrentBib}

\bibitem [\protect \citeauthoryear {%
{Prugniel}%
\ \BBA {} {Heraudeau}%
}{%
{Prugniel}%
\ \BBA {} {Heraudeau}%
}{%
{\protect \APACyear {1998}}%
}]{%
prugniel98}
\APACinsertmetastar {%
prugniel98}%
\begin{APACrefauthors}%
{Prugniel}, P.%
\BCBT {}\ \BBA {} {Heraudeau}, P.%
\end{APACrefauthors}%
\unskip\
\newblock
\APACrefYearMonthDay{1998}{{\APACmonth{03}}}{},
\newblock
\unskip
\newblock
\APACjournalVolNumPages{\aaps}{128}{}{299-308}.
\newblock
\begin{APACrefDOI} \doi{10.1051/aas:1998142} \end{APACrefDOI}
\PrintBackRefs{\CurrentBib}

\bibitem [\protect \citeauthoryear {%
{Pryor}%
\ \BBA {} {Meylan}%
}{%
{Pryor}%
\ \BBA {} {Meylan}%
}{%
{\protect \APACyear {1993}}%
}]{%
pryor93}
\APACinsertmetastar {%
pryor93}%
\begin{APACrefauthors}%
{Pryor}, C.%
\BCBT {}\ \BBA {} {Meylan}, G.%
\end{APACrefauthors}%
\unskip\
\newblock
\APACrefYearMonthDay{1993}{{\APACmonth{01}}}{},
\newblock
{\BBOQ}\APACrefatitle {{Velocity Dispersions for Galactic Globular Clusters}}
  {{Velocity Dispersions for Galactic Globular Clusters}}.{\BBCQ}
\newblock
\BIn{} S\BPBI G.~{Djorgovski}\ \BBA {} G.~{Meylan}\ (\BEDS), \APACrefbtitle
  {Structure and Dynamics of Globular Clusters} {Structure and Dynamics of
  Globular Clusters}\ \BVOL~50, \BPG~357.
\PrintBackRefs{\CurrentBib}

\bibitem [\protect \citeauthoryear {%
{Richtler}%
\ \protect \BOthers {.}}{%
{Richtler}%
\ \protect \BOthers {.}}{%
{\protect \APACyear {2004}}%
}]{%
richtler04}
\APACinsertmetastar {%
richtler04}%
\begin{APACrefauthors}%
{Richtler}, T.%
, {Dirsch}, B.%
, {Gebhardt}, K.%
\ et al.\end{APACrefauthors}%
\unskip\
\newblock
\APACrefYearMonthDay{2004}{{\APACmonth{04}}}{},
\newblock
\unskip
\newblock
\APACjournalVolNumPages{\aj}{127}{}{2094-2113}.
\newblock
\begin{APACrefDOI} \doi{10.1086/382721} \end{APACrefDOI}
\PrintBackRefs{\CurrentBib}

\bibitem [\protect \citeauthoryear {%
{Richtler}%
, {Salinas}%
, {Lane}%
, {Hilker}%
\BCBL {}\ \BBA {} {Schirmer}%
}{%
{Richtler}%
\ \protect \BOthers {.}}{%
{\protect \APACyear {2015}}%
}]{%
richtler15}
\APACinsertmetastar {%
richtler15}%
\begin{APACrefauthors}%
{Richtler}, T.%
, {Salinas}, R.%
, {Lane}, R\BPBI R.%
, {Hilker}, M.%
\BCBL {}\ \BBA {} {Schirmer}, M.%
\end{APACrefauthors}%
\unskip\
\newblock
\APACrefYearMonthDay{2015}{{\APACmonth{02}}}{},
\newblock
\unskip
\newblock
\APACjournalVolNumPages{\aap}{574}{}{A21}.
\newblock
\begin{APACrefDOI} \doi{10.1051/0004-6361/201424530} \end{APACrefDOI}
\PrintBackRefs{\CurrentBib}

\bibitem [\protect \citeauthoryear {%
{Richtler}%
\ \protect \BOthers {.}}{%
{Richtler}%
\ \protect \BOthers {.}}{%
{\protect \APACyear {2011}}%
}]{%
richtler11}
\APACinsertmetastar {%
richtler11}%
\begin{APACrefauthors}%
{Richtler}, T.%
, {Salinas}, R.%
, {Misgeld}, I.%
\ et al.\end{APACrefauthors}%
\unskip\
\newblock
\APACrefYearMonthDay{2011}{{\APACmonth{07}}}{},
\newblock
\unskip
\newblock
\APACjournalVolNumPages{\aap}{531}{}{A119+}.
\newblock
\begin{APACrefDOI} \doi{10.1051/0004-6361/201015948} \end{APACrefDOI}
\PrintBackRefs{\CurrentBib}

\bibitem [\protect \citeauthoryear {%
{Richtler}%
\ \protect \BOthers {.}}{%
{Richtler}%
\ \protect \BOthers {.}}{%
{\protect \APACyear {2008}}%
}]{%
richtler08}
\APACinsertmetastar {%
richtler08}%
\begin{APACrefauthors}%
{Richtler}, T.%
, {Schuberth}, Y.%
, {Hilker}, M.%
, {Dirsch}, B.%
, {Bassino}, L.%
\BCBL {}\ \BBA {} {Romanowsky}, A\BPBI J.%
\end{APACrefauthors}%
\unskip\
\newblock
\APACrefYearMonthDay{2008}{{\APACmonth{02}}}{},
\newblock
\unskip
\newblock
\APACjournalVolNumPages{\aap}{478}{}{L23-L26}.
\newblock
\begin{APACrefDOI} \doi{10.1051/0004-6361:20078539} \end{APACrefDOI}
\PrintBackRefs{\CurrentBib}

\bibitem [\protect \citeauthoryear {%
{Romanowsky}%
\ \protect \BOthers {.}}{%
{Romanowsky}%
\ \protect \BOthers {.}}{%
{\protect \APACyear {2003}}%
}]{%
romanowsky03}
\APACinsertmetastar {%
romanowsky03}%
\begin{APACrefauthors}%
{Romanowsky}, A\BPBI J.%
, {Douglas}, N\BPBI G.%
, {Arnaboldi}, M.%
\ et al.\end{APACrefauthors}%
\unskip\
\newblock
\APACrefYearMonthDay{2003}{{\APACmonth{09}}}{},
\newblock
\unskip
\newblock
\APACjournalVolNumPages{Science}{301}{}{1696-1698}.
\newblock
\begin{APACrefDOI} \doi{10.1126/science.1087441} \end{APACrefDOI}
\PrintBackRefs{\CurrentBib}

\bibitem [\protect \citeauthoryear {%
{Sales}%
\ \protect \BOthers {.}}{%
{Sales}%
\ \protect \BOthers {.}}{%
{\protect \APACyear {2017}}%
}]{%
sales17}
\APACinsertmetastar {%
sales17}%
\begin{APACrefauthors}%
{Sales}, L\BPBI V.%
, {Navarro}, J\BPBI F.%
, {Oman}, K.%
\ et al.\end{APACrefauthors}%
\unskip\
\newblock
\APACrefYearMonthDay{2017}{{\APACmonth{01}}}{},
\newblock
\unskip
\newblock
\APACjournalVolNumPages{\mnras}{464}{}{2419-2428}.
\newblock
\begin{APACrefDOI} \doi{10.1093/mnras/stw2461} \end{APACrefDOI}
\PrintBackRefs{\CurrentBib}

\bibitem [\protect \citeauthoryear {%
{Salinas}%
, {Richtler}%
, {Bassino}%
, {Romanowsky}%
\BCBL {}\ \BBA {} {Schuberth}%
}{%
{Salinas}%
\ \protect \BOthers {.}}{%
{\protect \APACyear {2012}}%
}]{%
salinas12}
\APACinsertmetastar {%
salinas12}%
\begin{APACrefauthors}%
{Salinas}, R.%
, {Richtler}, T.%
, {Bassino}, L\BPBI P.%
, {Romanowsky}, A\BPBI J.%
\BCBL {}\ \BBA {} {Schuberth}, Y.%
\end{APACrefauthors}%
\unskip\
\newblock
\APACrefYearMonthDay{2012}{{\APACmonth{02}}}{},
\newblock
\unskip
\newblock
\APACjournalVolNumPages{\aap}{538}{}{A87}.
\newblock
\begin{APACrefDOI} \doi{10.1051/0004-6361/201116517} \end{APACrefDOI}
\PrintBackRefs{\CurrentBib}

\bibitem [\protect \citeauthoryear {%
{Sanders}%
}{%
{Sanders}%
}{%
{\protect \APACyear {2003}}%
}]{%
sanders03}
\APACinsertmetastar {%
sanders03}%
\begin{APACrefauthors}%
{Sanders}, R\BPBI H.%
\end{APACrefauthors}%
\unskip\
\newblock
\APACrefYearMonthDay{2003}{{\APACmonth{07}}}{},
\newblock
\unskip
\newblock
\APACjournalVolNumPages{\mnras}{342}{}{901-908}.
\newblock
\begin{APACrefDOI} \doi{10.1046/j.1365-8711.2003.06596.x} \end{APACrefDOI}
\PrintBackRefs{\CurrentBib}

\bibitem [\protect \citeauthoryear {%
{Schuberth}%
\ \protect \BOthers {.}}{%
{Schuberth}%
\ \protect \BOthers {.}}{%
{\protect \APACyear {2006}}%
}]{%
schuberth06}
\APACinsertmetastar {%
schuberth06}%
\begin{APACrefauthors}%
{Schuberth}, Y.%
, {Richtler}, T.%
, {Dirsch}, B.%
, {Hilker}, M.%
, {Larsen}, S\BPBI S.%
, {Kissler-Patig}, M.%
\BCBL {}\ \BBA {} {Mebold}, U.%
\end{APACrefauthors}%
\unskip\
\newblock
\APACrefYearMonthDay{2006}{{\APACmonth{11}}}{},
\newblock
\unskip
\newblock
\APACjournalVolNumPages{\aap}{459}{}{391-406}.
\newblock
\begin{APACrefDOI} \doi{10.1051/0004-6361:20053134} \end{APACrefDOI}
\PrintBackRefs{\CurrentBib}

\bibitem [\protect \citeauthoryear {%
{Schuberth}%
\ \protect \BOthers {.}}{%
{Schuberth}%
\ \protect \BOthers {.}}{%
{\protect \APACyear {2010}}%
}]{%
schuberth10}
\APACinsertmetastar {%
schuberth10}%
\begin{APACrefauthors}%
{Schuberth}, Y.%
, {Richtler}, T.%
, {Hilker}, M.%
, {Dirsch}, B.%
, {Bassino}, L\BPBI P.%
, {Romanowsky}, A\BPBI J.%
\BCBL {}\ \BBA {} {Infante}, L.%
\end{APACrefauthors}%
\unskip\
\newblock
\APACrefYearMonthDay{2010}{{\APACmonth{04}}}{},
\newblock
\unskip
\newblock
\APACjournalVolNumPages{\aap}{513}{}{A52+}.
\newblock
\begin{APACrefDOI} \doi{10.1051/0004-6361/200912482} \end{APACrefDOI}
\PrintBackRefs{\CurrentBib}

\bibitem [\protect \citeauthoryear {%
{Schuberth}%
\ \protect \BOthers {.}}{%
{Schuberth}%
\ \protect \BOthers {.}}{%
{\protect \APACyear {2012}}%
}]{%
schuberth12}
\APACinsertmetastar {%
schuberth12}%
\begin{APACrefauthors}%
{Schuberth}, Y.%
, {Richtler}, T.%
, {Hilker}, M.%
, {Salinas}, R.%
, {Dirsch}, B.%
\BCBL {}\ \BBA {} {Larsen}, S\BPBI S.%
\end{APACrefauthors}%
\unskip\
\newblock
\APACrefYearMonthDay{2012}{{\APACmonth{08}}}{},
\newblock
\unskip
\newblock
\APACjournalVolNumPages{\aap}{544}{}{A115}.
\newblock
\begin{APACrefDOI} \doi{10.1051/0004-6361/201015038} \end{APACrefDOI}
\PrintBackRefs{\CurrentBib}

\bibitem [\protect \citeauthoryear {%
{Serra}%
, {Oosterloo}%
, {Cappellari}%
, {den Heijer}%
\BCBL {}\ \BBA {} {J{\'o}zsa}%
}{%
{Serra}%
\ \protect \BOthers {.}}{%
{\protect \APACyear {2016}}%
}]{%
serra16}
\APACinsertmetastar {%
serra16}%
\begin{APACrefauthors}%
{Serra}, P.%
, {Oosterloo}, T.%
, {Cappellari}, M.%
, {den Heijer}, M.%
\BCBL {}\ \BBA {} {J{\'o}zsa}, G\BPBI I\BPBI G.%
\end{APACrefauthors}%
\unskip\
\newblock
\APACrefYearMonthDay{2016}{{\APACmonth{08}}}{},
\newblock
\unskip
\newblock
\APACjournalVolNumPages{\mnras}{460}{}{1382-1389}.
\newblock
\begin{APACrefDOI} \doi{10.1093/mnras/stw1010} \end{APACrefDOI}
\PrintBackRefs{\CurrentBib}

\bibitem [\protect \citeauthoryear {%
{Silverwood}%
, {Weniger}%
, {Scott}%
\BCBL {}\ \BBA {} {Bertone}%
}{%
{Silverwood}%
\ \protect \BOthers {.}}{%
{\protect \APACyear {2015}}%
}]{%
silverwood15}
\APACinsertmetastar {%
silverwood15}%
\begin{APACrefauthors}%
{Silverwood}, H.%
, {Weniger}, C.%
, {Scott}, P.%
\BCBL {}\ \BBA {} {Bertone}, G.%
\end{APACrefauthors}%
\unskip\
\newblock
\APACrefYearMonthDay{2015}{{\APACmonth{03}}}{},
\newblock
\unskip
\newblock
\APACjournalVolNumPages{\jcap}{3}{}{055}.
\newblock
\begin{APACrefDOI} \doi{10.1088/1475-7516/2015/03/055} \end{APACrefDOI}
\PrintBackRefs{\CurrentBib}

\bibitem [\protect \citeauthoryear {%
{Skordis}%
\ \BBA {} {Z{\l}o{\'s}nik}%
}{%
{Skordis}%
\ \BBA {} {Z{\l}o{\'s}nik}%
}{%
{\protect \APACyear {2021}}%
}]{%
skordis21}
\APACinsertmetastar {%
skordis21}%
\begin{APACrefauthors}%
{Skordis}, C.%
\BCBT {}\ \BBA {} {Z{\l}o{\'s}nik}, T.%
\end{APACrefauthors}%
\unskip\
\newblock
\APACrefYearMonthDay{2021}{{\APACmonth{10}}}{},
\newblock
\unskip
\newblock
\APACjournalVolNumPages{\prl}{127}{16}{161302}.
\newblock
\begin{APACrefDOI} \doi{10.1103/PhysRevLett.127.161302} \end{APACrefDOI}
\PrintBackRefs{\CurrentBib}

\bibitem [\protect \citeauthoryear {%
{Stark}%
, {McGaugh}%
\BCBL {}\ \BBA {} {Swaters}%
}{%
{Stark}%
\ \protect \BOthers {.}}{%
{\protect \APACyear {2009}}%
}]{%
stark09}
\APACinsertmetastar {%
stark09}%
\begin{APACrefauthors}%
{Stark}, D\BPBI V.%
, {McGaugh}, S\BPBI S.%
\BCBL {}\ \BBA {} {Swaters}, R\BPBI A.%
\end{APACrefauthors}%
\unskip\
\newblock
\APACrefYearMonthDay{2009}{{\APACmonth{08}}}{},
\newblock
\unskip
\newblock
\APACjournalVolNumPages{\aj}{138}{}{392-401}.
\newblock
\begin{APACrefDOI} \doi{10.1088/0004-6256/138/2/392} \end{APACrefDOI}
\PrintBackRefs{\CurrentBib}

\bibitem [\protect \citeauthoryear {%
{Strominger}%
\ \BBA {} {Vafa}%
}{%
{Strominger}%
\ \BBA {} {Vafa}%
}{%
{\protect \APACyear {1996}}%
}]{%
strominger96}
\APACinsertmetastar {%
strominger96}%
\begin{APACrefauthors}%
{Strominger}, A.%
\BCBT {}\ \BBA {} {Vafa}, C.%
\end{APACrefauthors}%
\unskip\
\newblock
\APACrefYearMonthDay{1996}{{\APACmonth{02}}}{},
\newblock
\unskip
\newblock
\APACjournalVolNumPages{Physics Letters B}{379}{}{99-104}.
\newblock
\begin{APACrefDOI} \doi{10.1016/0370-2693(96)00345-0} \end{APACrefDOI}
\PrintBackRefs{\CurrentBib}

\bibitem [\protect \citeauthoryear {%
{Trachternach}%
, {de Blok}%
, {McGaugh}%
, {van der Hulst}%
\BCBL {}\ \BBA {} {Dettmar}%
}{%
{Trachternach}%
\ \protect \BOthers {.}}{%
{\protect \APACyear {2009}}%
}]{%
trachternach09}
\APACinsertmetastar {%
trachternach09}%
\begin{APACrefauthors}%
{Trachternach}, C.%
, {de Blok}, W\BPBI J\BPBI G.%
, {McGaugh}, S\BPBI S.%
, {van der Hulst}, J\BPBI M.%
\BCBL {}\ \BBA {} {Dettmar}, R.%
\end{APACrefauthors}%
\unskip\
\newblock
\APACrefYearMonthDay{2009}{{\APACmonth{10}}}{},
\newblock
\unskip
\newblock
\APACjournalVolNumPages{\aap}{505}{}{577-587}.
\newblock
\begin{APACrefDOI} \doi{10.1051/0004-6361/200811136} \end{APACrefDOI}
\PrintBackRefs{\CurrentBib}

\bibitem [\protect \citeauthoryear {%
{van der Marel}%
\ \BBA {} {Franx}%
}{%
{van der Marel}%
\ \BBA {} {Franx}%
}{%
{\protect \APACyear {1993}}%
}]{%
vandermarel93}
\APACinsertmetastar {%
vandermarel93}%
\begin{APACrefauthors}%
{van der Marel}, R\BPBI P.%
\BCBT {}\ \BBA {} {Franx}, M.%
\end{APACrefauthors}%
\unskip\
\newblock
\APACrefYearMonthDay{1993}{{\APACmonth{04}}}{},
\newblock
\unskip
\newblock
\APACjournalVolNumPages{\apj}{407}{}{525}.
\newblock
\begin{APACrefDOI} \doi{10.1086/172534} \end{APACrefDOI}
\PrintBackRefs{\CurrentBib}

\bibitem [\protect \citeauthoryear {%
{Vazdekis}%
\ \protect \BOthers {.}}{%
{Vazdekis}%
\ \protect \BOthers {.}}{%
{\protect \APACyear {2010}}%
}]{%
vazdekis10}
\APACinsertmetastar {%
vazdekis10}%
\begin{APACrefauthors}%
{Vazdekis}, A.%
, {S{\'a}nchez-Bl{\'a}zquez}, P.%
, {Falc{\'o}n-Barroso}, J.%
\ et al.\end{APACrefauthors}%
\unskip\
\newblock
\APACrefYearMonthDay{2010}{{\APACmonth{06}}}{},
\newblock
\unskip
\newblock
\APACjournalVolNumPages{\mnras}{404}{4}{1639-1671}.
\newblock
\begin{APACrefDOI} \doi{10.1111/j.1365-2966.2010.16407.x} \end{APACrefDOI}
\PrintBackRefs{\CurrentBib}

\bibitem [\protect \citeauthoryear {%
{Verlinde}%
}{%
{Verlinde}%
}{%
{\protect \APACyear {2011}}%
}]{%
verlinde11}
\APACinsertmetastar {%
verlinde11}%
\begin{APACrefauthors}%
{Verlinde}, E.%
\end{APACrefauthors}%
\unskip\
\newblock
\APACrefYearMonthDay{2011}{{\APACmonth{04}}}{},
\newblock
\unskip
\newblock
\APACjournalVolNumPages{Journal of High Energy Physics}{4}{}{29}.
\newblock
\begin{APACrefDOI} \doi{10.1007/JHEP04(2011)029} \end{APACrefDOI}
\PrintBackRefs{\CurrentBib}

\bibitem [\protect \citeauthoryear {%
{Verlinde}%
}{%
{Verlinde}%
}{%
{\protect \APACyear {2017}}%
}]{%
verlinde17}
\APACinsertmetastar {%
verlinde17}%
\begin{APACrefauthors}%
{Verlinde}, E.%
\end{APACrefauthors}%
\unskip\
\newblock
\APACrefYearMonthDay{2017}{{\APACmonth{05}}}{},
\newblock
\unskip
\newblock
\APACjournalVolNumPages{SciPost Physics}{2}{}{016}.
\newblock
\begin{APACrefDOI} \doi{10.21468/SciPostPhys.2.3.016} \end{APACrefDOI}
\PrintBackRefs{\CurrentBib}

\bibitem [\protect \citeauthoryear {%
{Watkins}%
, {Evans}%
\BCBL {}\ \BBA {} {An}%
}{%
{Watkins}%
\ \protect \BOthers {.}}{%
{\protect \APACyear {2010}}%
}]{%
watkins10}
\APACinsertmetastar {%
watkins10}%
\begin{APACrefauthors}%
{Watkins}, L\BPBI L.%
, {Evans}, N\BPBI W.%
\BCBL {}\ \BBA {} {An}, J\BPBI H.%
\end{APACrefauthors}%
\unskip\
\newblock
\APACrefYearMonthDay{2010}{{\APACmonth{07}}}{},
\newblock
\unskip
\newblock
\APACjournalVolNumPages{\mnras}{406}{}{264-278}.
\newblock
\begin{APACrefDOI} \doi{10.1111/j.1365-2966.2010.16708.x} \end{APACrefDOI}
\PrintBackRefs{\CurrentBib}

\bibitem [\protect \citeauthoryear {%
{Weijmans}%
\ \protect \BOthers {.}}{%
{Weijmans}%
\ \protect \BOthers {.}}{%
{\protect \APACyear {2008}}%
}]{%
weijmans08}
\APACinsertmetastar {%
weijmans08}%
\begin{APACrefauthors}%
{Weijmans}, A\BHBI M.%
, {Krajnovi{\'c}}, D.%
, {van de Ven}, G.%
, {Oosterloo}, T\BPBI A.%
, {Morganti}, R.%
\BCBL {}\ \BBA {} {de Zeeuw}, P\BPBI T.%
\end{APACrefauthors}%
\unskip\
\newblock
\APACrefYearMonthDay{2008}{{\APACmonth{02}}}{},
\newblock
\unskip
\newblock
\APACjournalVolNumPages{\mnras}{383}{}{1343-1358}.
\newblock
\begin{APACrefDOI} \doi{10.1111/j.1365-2966.2007.12680.x} \end{APACrefDOI}
\PrintBackRefs{\CurrentBib}

\bibitem [\protect \citeauthoryear {%
{White}%
}{%
{White}%
}{%
{\protect \APACyear {1981}}%
}]{%
white81}
\APACinsertmetastar {%
white81}%
\begin{APACrefauthors}%
{White}, S\BPBI D\BPBI M.%
\end{APACrefauthors}%
\unskip\
\newblock
\APACrefYearMonthDay{1981}{{\APACmonth{06}}}{},
\newblock
\unskip
\newblock
\APACjournalVolNumPages{\mnras}{195}{}{1037-1056}.
\newblock
\begin{APACrefDOI} \doi{10.1093/mnras/195.4.1037} \end{APACrefDOI}
\PrintBackRefs{\CurrentBib}

\bibitem [\protect \citeauthoryear {%
{Willmer}%
}{%
{Willmer}%
}{%
{\protect \APACyear {2018}}%
}]{%
willmer18}
\APACinsertmetastar {%
willmer18}%
\begin{APACrefauthors}%
{Willmer}, C\BPBI N\BPBI A.%
\end{APACrefauthors}%
\unskip\
\newblock
\APACrefYearMonthDay{2018}{{\APACmonth{06}}}{},
\newblock
\unskip
\newblock
\APACjournalVolNumPages{\apjs}{236}{2}{47}.
\newblock
\begin{APACrefDOI} \doi{10.3847/1538-4365/aabfdf} \end{APACrefDOI}
\PrintBackRefs{\CurrentBib}

\end{thebibliography}
\appendix
\section{Tables}
Table A1 lists the sample of slits of the galaxy light,  for which radial velocities  and velocity dispersions could be measured.   The columns are the identification (the first figure is the mask number,
followed by the slit number), the coordinates (J2000),   the projected galactocentric distance in kpc, the
heliocentric radial velocity, its uncertainty,  the velocity dispersion and its uncertainty.  Table A3 identifies the stars in our slit sample. 

\begin{table*}
\centering
%\resizebox{0.45\textwidth}{!}{
\begin{tabular}{cccccccccccc}
\hline
\hline
Id  &  RA(J2000)  & Dec(J2000) &  radius[kpc] & $V_r$ &  error & $\sigma_{LOS}$ & error & $\sigma_{LOS}$& error  & metallicity & log(age)  \\
     &                      &                    &                    &            &           &    4 moments       &             &    2 moments       &               \\          
\hline
116  & 225.230398       & -7.458710 & 1.002 &  1965 & 7&  172 & 11 &  188 &   11 & -0.01 & 1.14 \\ 
19   & 225.231242       & -7.460070 & 1.416 &   1934 & 6 & 151 &  10 & 163 & 10   & -0.06 & 1.14 \\
115  & 225.230784       & -7.455860 & 0.863 & 1974& 4 & 191 & 8  & 210  &   8  & 0.0 &  1.16\\ 
118  & 225.233402       & -7.459140 &1.214 & 1910 & 6 & 159 & 10 & 181  &  7.1  & -0.06 & 1.15  \\
114  & 225.231142       & -7.453820 &1.720 &  1919 & 6 & 153 & 11 & 161   &    5  & -0.08 & 1.15  \\ 
14  & 225.233331       & -7.455870& 0.984 &   1903 & 4.2 & 180 & 9 & 196   &   5   & -0.05 & 1.16 \\ 
15  & 225.231929       & -7.451300 & 2.915 &   1924&  4.6& 138 &  10 &  151 &  5  &  & \\ 
%117  & 225.238366       & -7.450610 &4.519  &  1922 & 9 &161    &   13  & 167 & 9   \\
122  & 225.229840       & -7.464870 &3.846  & 1935 & 9 & 117 &   13 & 124 & 7 & -0.34 & 1.1  \\ 
221   & 225.232859       & -7.455040 &1.190  &  1931 &  6 &172   &   10 & 191 & 6  & -0.01 & 1.16\\ 
227  & 225.227022       & -7.458690 &2.445 &  1973 &  7 & 134 & 11 & 144 &   7.5 &   -0.13 & 1.16 \\
226   & 225.236506       & -7.452520 &3.231 &  1905 &  6  &108  & 10 & 115 &   8  &  -0.24 & 1.16 \\
%220   & 225.223775       & -7.455320 & 4.036 &  1995 &  9 & 123 &    15  &  94 & 13\\
232   & 225.239654       & -7.458670 &3.831 &  1900 & 6.8 &156 &   8 &  151 & 16  & -0.38 & 1.16\\
%230   & 225.229840       & -7.464870 &3.846 &  1960 & 7.6 &126  &    14 & 134 & 8 \\
219  & 225.223761       & -7.459020 &3.824 &  1998 & 8.8 & 105 &  15  & 109 & 10 & -0.2 & 1.16\\
%223  & 225.239096       & -7.466630 &5.746 &  1902 & 11& 162 &   12  & 162 & 13  \\    star!
\hline
\end{tabular}
%}
\label{tab:galaxyslits}
\caption{Sample data for our galaxy spectra. The columns are: identifier (first number is the mask number, second number the slit number), coordinates, projected galactocentric distance, radial velocities and their uncertainties,  velocity dispersion and
their uncertainties. The coordinates
are the frame WCS coordinates.}
\end{table*}

Table A2 lists the sample of GCs  for which radial velocities could be measured.  These are  37 measurements for 29 globular clusters  and 19 stars.  The columns are the identification (the first digit is the mask number,
followed by the slit number), the coordinates (J2000),   the
heliocentric radial velocity, its uncertainty,  the R-magnitude, the colour C-R.  
\begin{table*}
\centering
\resizebox{0.75\textwidth}{!}{
\begin{tabular}{rcccccccccccc}
\hline
\hline
Id  &  RA(J2000)  & Dec(J2000) &  $v_r$ (N1396)  &  error & $v_r$ (GC) & error&  $v_r(final)$ & error & C-R & R &  double  \\
\hline
%n1316\_gc00250 & 3:22:13.73 & $-$37:11:56.5 &  23.30 &  1.20 &  2020.0 &  61.0 \\
%n1316\_gc00280 & 3:22:23.21 & $-$37:08:26.0 &  22.05 &  5.07 &  1656.1 &  42.9 \\
%n1316\_gc00285 & 3:22:25.89 & $-$37:12:02.1 &  23.50 &  1.32 &  1869.2 &  56.7 \\
%n1316\_gc00286 & 3:22:26.02 & $-$37:05:17.1 &  23.30 &  1.33 &  1427.0 &  49.0 \\
   15  & 225.23193 & -7.45130 & 1942&  29 &     &   & 1942&  29 &     &     &  \\
 16  & 225.24197 & -7.44785 & 1978&  43 & 1909 &  41& 1944&  42 & 1.42 & 22.81&  \\
 18  & 225.24789 & -7.43559 & 1891&  25 & 1873 &  25& 1882&  25& 1.64 & 20.34& 213 \\
 111  & 225.23985 & -7.44285 & 1968&  41 & 1940 &  23& 1954&  33& 1.53& 21.48& \\ 
 117  & 225.23837 & -7.45061 & 1974&  31 & 1955 &  37& 1964&  34& 1.47& 21.81 & 216 \\
 119  & 225.22414 & -7.45902  & 2071& 19 &2053 &  19& 2062&  19& 1.70& 21.89& 219 \\
 120  & 225.21091 & -7.45582 & 2010&  71 & 1990 &  47& 2000&  60& 1.88& 21.49&  \\
 122  & 225.22984 & -7.46487 & 1956&  40 & 2008 &  38& 1982&  39& 1.41& 22.51& 230\\
 123  & 225.23465 & -7.47690 & 1949&  27 &1936 &  27& 1942&  27& 1.99& 22.16& 235\\
 124  & 225.21896 & -7.46620 & 1943&  30 & 1885 &  23& 1914&  27& 1.77& 22.55& 238\\
 125  & 225.22070 & -7.47564 & 2109&  20& 2101 &  16& 2105&  18& 1.43& 21.81&  \\
 126  & 225.23170 & -7.47320 & 1880&  34& 1889 &  32& 1884&  33& 1.17 &22.10&  \\
 127  & 225.21230 & -7.47584 & 1933&  67& 1924 &  57& 1928&  62& 1.33& 22.50&  \\
 128  & 225.23160 & -7.48925 & 1791&  &     &  & 1791&  99& 1.16& 22.71& 241\\
 28  & 225.23492 & -7.43592 & 1545&  90& 1475 &  73 &1510&  82 &1.26& 22.62&  \\
 212  & 225.25096 & -7.43856 &2126&  74 &    &   & 2126& 74 & 1.29& 22.97&  \\
 213  & 225.24789 & -7.43559 & 1934&  21 &1926 &  28& 1930&  25 &1.64 & 20.34& 314\\
 216  & 225.23846 & -7.45039 & 1932&  55 &1904 &  58& 1918&  57& 1.47& 21.82&  \\
 217  & 225.23776 & -7.46174 & 2024&  35& 1967 &  28& 1996&  32& 1.65& 21.26&  \\
 219  & 225.22375 & -7.45902 &  2174&  64& 2127 &  52& 2150&  58 &1.69 &21.89&  \\
 220  & 225.22377 & -7.45532 & 2103&  78 &2114 &  38& 2108&  61&    &    &  \\
 222  & 225.23132 & -7.44329 & 2104  &77    & &     & 2104 &     77  &  1.28& 22.43& \\ 
 224  & 225.23109 & -7.44894 & 1742&  68& 1829 & 150& 1786& 116& 1.06& 22.75&  \\
 229  & 225.24104 & -7.45586 &1852&  37& 1820 &  25 & 1836&  32& 1.56& 22.17&  \\
 230  & 225.22984 & -7.46487 & 2067&  51& 2059 &  52& 2063&  52& 1.41& 22.51&  \\
 233  & 225.23882 & -7.46956 &1804&  34& 1760 &  23& 1782&  29& 1.38& 22.82&  \\
 234  & 225.23547 & -7.48179 & 1946&  70 &  &  & 1946&  70& 1.48 &22.04&  \\
 235  & 225.23465 & -7.47690 & 2073&  34 & 2002 &  25 & 2038&  30& 1.99& 22.16&  \\
 236  & 225.23518 & -7.48353 &1753&  41 &1742 &  50 &1748&  46& 1.03& 22.73&  \\
 239  & 225.24402 & -7.47851 & 1926&  39& 1921 &  45& 1924&  42& 1.61 &21.85&  \\
 240  & 225.22774 & -7.48521 & 1766&  71& 1740 &  73& 1753&  72&    &    &  \\
 243  & 225.21761 & -7.48575 & 1835&  38&     &  99& 1417&  75& 1.25& 22.73&  \\
 244  & 225.23170 & -7.49690 & 1985&  44 &1893 &  42 &1939&  43& 1.05& 23.27&  \\
 314  & 225.24779 & -7.43589 & 1948&  39& 1915 &  38& 1932&  39& 1.64 & 20.34&  \\
 315  & 225.25200 & -7.41978 & 2027&  56& 1925 &  28& 1976&  44& 1.32& 21.67&  \\
 317  & 225.24181 & -7.42524 & 1981&  50 & 2019 &  49& 2000&  50 &1.34 &21.91&  \\
 322  & 225.23480 & -7.43623 & 1790&  60 &1739 &  45& 1764&  53& 1.26 &22.62&  \\
 \hline
\end{tabular}
}
\label{tab:gcsample}
\caption{Identification (mask number and slit number), coordinates, radial velocities, C-R colours, and R-magnitude, for the globular cluster sample. The last column identifies
double measurements. The coordinates
are the frame WCS coordinates and for identification purposes only. The photometry is adopted from the catalogue of \citet{lane13}.}
\end{table*}

\begin{table*}
\centering
\resizebox{0.75\textwidth}{!}{
\begin{tabular}{rcccccccccccc}
\hline
\hline
Id  &  RA(J2000)  & Dec(J2000) &  $v_r$ (N1396)  &  error & $v_r$ (GC) & error&  $v_r(final)$ & error & R & C-R &  double  \\
\hline

 11 &225.24913 &-7.41951 &-104 &37 &-124 &33 &-114 &35 &1.58 &21.56 &22 \\
  12 &225.22849 &-7.41639 &1664 &33 &     &   &     &   &     &      & dwarf \\
  13 &225.26917 &-7.43612 &9 &28 &-14 &36 &-2 &32 &1.63 &20.95 &21 \\
  17 &225.23570 &-7.43727 &217 &22 &194 &28 &206 &25 &1.26 &20.66 &\\
 131 &225.23338 &-7.50151 &103 &25 &89 &30 &96 &28 &1.03 &22.05 & \\
  21 &225.26917 &-7.43612 &5 &27 &-5 &30 &0 &29 &1.63 &20.95 & \\
  22 &225.24913 &-7.41951 &-135 &71 &-118 &44 &-126 &59 &1.58 &21.56& \\
  23 &225.24194 &-7.42020 &104 &66 &105 &63 &104 &65 &1.18 &19.06 &318 \\
 211 &225.27092 &-7.45106 &-208 &27 &-234 &32 &-221 &30 &1.62 &20.93 & \\
 215 &225.26537 &-7.43833 &-112 &21 &-115 &24 &-114 &23 &1.36 &21.61 &316 \\
 223 &225.23909 &-7.46663 &-15 &48 &-32 &45 &-24 &47 &1.41 &23.56 &\\
 225 &225.23969 &-7.45997 &79 &24 &70 &28 &74 &26 &1.33 &21.11& \\
 237 &225.21407 &-7.47114 &-27 &33 &-42 &22 &-34 &28 &1.52 &22.57 & \\
 242 &225.25816 &-7.50252 &114 &37 &99 &54 &106 &46 &1.60 &19.51 & \\
 246 &225.21674 &-7.48863 &-39 &23 &-47 &19 &-43 &21 & & & \\
 316 &225.26523 &-7.43866 &-20 &104 &-37 &120 &-28 &112 &1.36 &21.61 &\\
 318 &225.24184 &-7.42052 &118 &58 &84 &52 &101 &55 &1.18 &19.06 &\\
 319 &225.26250 &-7.41579 &92 &28 &62 &-36 &77 &32 &1.24 &17.38 &\\
 320 &225.23843 &-7.41054 &112 &45 &28 &38 &70 &42 &1.04 &22.05 &\\
 \hline
\end{tabular}
}
\label{tab:stars} 
\caption{Identification(mask number and slit number), coordinates, radial velocities, C-R colours, R-magnitudes  for the stars in our sample. Object 12 is the dwarf companion. The coordinates
are for identification purposes only. The photometry is adopted from the catalogue of \citet{lane13}.}
\end{table*}

%\begin{table}
%\centering
%\begin{tabular}{ccccccc}
%\hline\hline
%n1316\_gc00250 & 3:22:13.73 & $-$37:11:56.5 &  23.30 &  1.20 &  2020.0 &  61.0 \\
%n1316\_gc00280 & 3:22:23.21 & $-$37:08:26.0 &  22.05 &  5.07 &  1656.1 &  42.9 \\
%n1316\_gc00285 & 3:22:25.89 & $-$37:12:02.1 &  23.50 &  1.32 &  1869.2 &  56.7 \\
%n1316\_gc00286 & 3:22:26.02 & $-$37:05:17.1 &  23.30 &  1.33 &  1427.0 &  49.0 \\
%\hline\hline
%\normalsize
%\end{tabular}
%\label{gcsample}
%\caption{Identification and radial velocities of stars, galaxies, quasars}
%\end{table}

%\section{Sample spectra}
%As an illustration, we show some of our brighter spectra. 

%\section{Spectra}

\begin{figure*}[]
\begin{center}
\includegraphics[width=0.8\textwidth]{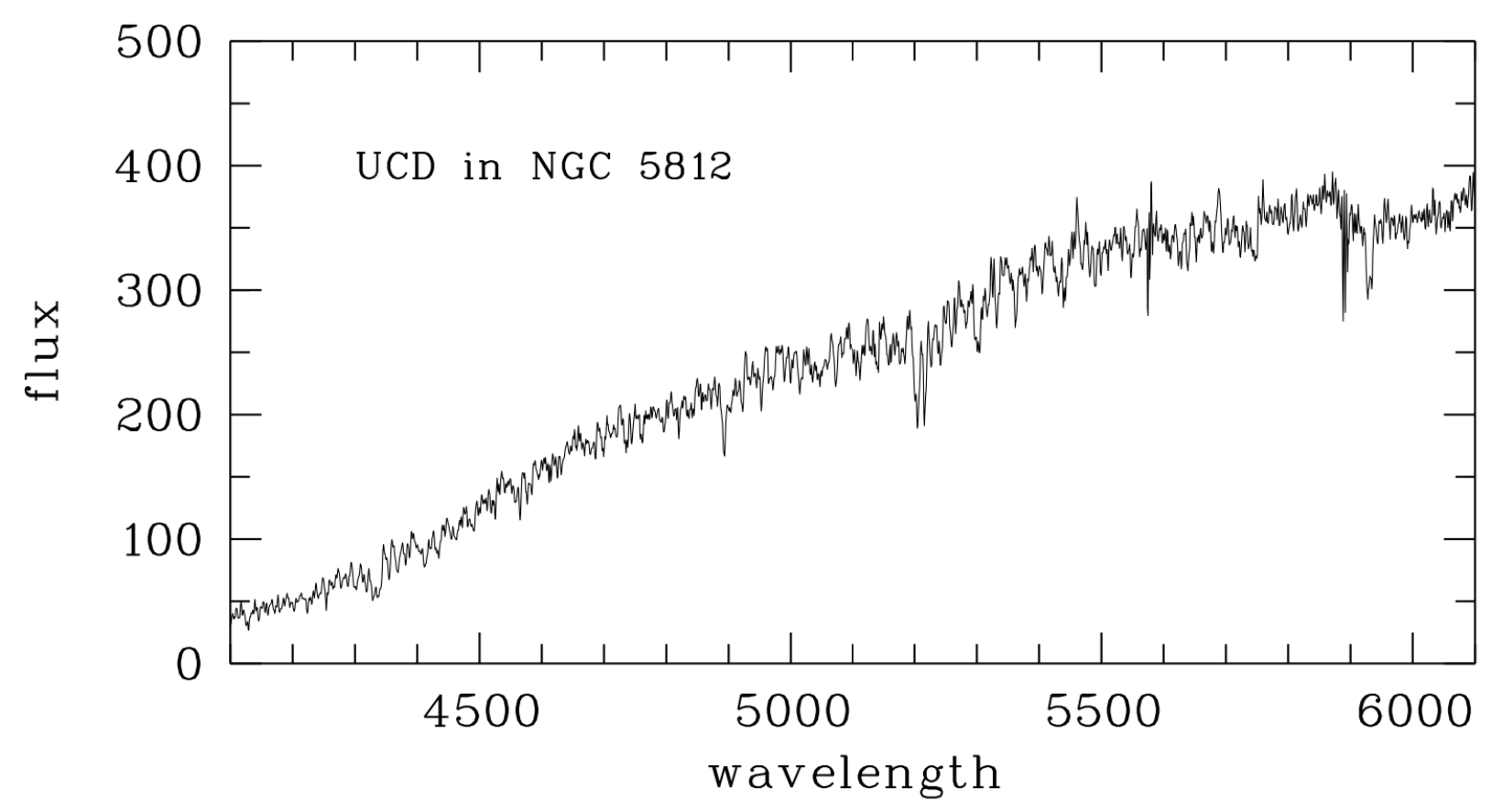}
\caption{This globular cluster has the absolute magnitude $M_R$ = -12.2. It is old and moderately metal-poor (see section \ref{sec:UCD}). We estimate its mass to  $1.6\times10^7 M_\odot$.}
\label{fig:ucd}
\end{center}
\end{figure*}

%\sections{todolist}
%Fig2.\\
%skysubtraction of galaxy slits \\
%mentioning separation stars and clusters\\
%context of specific frequency\\
%check the numbering\\

%\bibliographystyle{aa}
%{N5812new}
%\bibliography{N1316.bib}
\end{document}